\documentclass[10pt]{iopart}
\usepackage{iopams}
\usepackage{graphicx}
\usepackage{hyperref}
\newcommand{\mmm}{\mathbf m}

\begin{document}
\title[Mechanics of a domain wall]{Mechanics of a ferromagnetic domain wall}
\author{Se Kwon Kim$^1$ and Oleg Tchernyshyov$^2$}
\address{$^1$ Department of Physics, Korea Advanced Institute of Science and Technology, Daejeon 34141, Republic of Korea}
\address{$^2$ William H Miller III Department of Physics and Astronomy and Institute for Quantum Matter, Johns Hopkins University, Baltimore, MD 21218, USA}
\ead{$^1$sekwonkim@kaist.ac.kr, $^2$olegt@jhu.edu}
\begin{abstract}
This paper gives a pedagogical introduction to the mechanics of ferromagnetic solitons. We start with the dynamics of a single spin and develop all the tools required for the description of the dynamics of solitons in a ferromagnet. 
\end{abstract}
\submitto{\JPCM}
\maketitle
\tableofcontents

\section{Introduction}

This article aims to provide a self-contained introduction to theory of ferromagnetic solitons. The subject has been actively researched for more than 50 years and there are a number of excellent reviews and books \cite{Kosevich1990, bar2006dynamics}. The last decades have seen new developments in the field such as the use of collective coordinates to describe the dynamics of solitons \cite{Tretiakov2008, ZangPRL2011, ThiavilleEPL2012, TataraPRB2020} as well as in technological applications such as soliton-based racetrack memory~\cite{FertNN2013, ParkinNN2015} 

A ferromagnet contains a large number of atomic magnetic dipoles that tend to line up with one another at sufficiently low temperatures. The interaction most commonly responsible for the parallel alignment is Heisenberg's exchange, which represents a quantum-mechanical effect related to the fermionic statistics of electrons. One of the most frequently used models of a ferromagnet is the Heisenberg model with quantum spins of length $S=\hbar/2$ on a lattice with nearest-neighbor exchange interactions. The ground state has all spins pointing in the same direction. Elementary excitations are magnons---quasiparticles carrying spin $S_z = -\hbar$ along the direction of magnetization. A magnon mode in a coherent state with a large amplitude represents a classical wave of magnetization (a spin wave)~\cite{Altland-Simons}. 

This widespread approach, starting with a quantum model of spins on a lattice, is inconvenient for our purposes. Although it is suitable for obtaining weakly excited states such as magnons, the treatment of topological solitons at the quantum level and on a lattice is too complicated. Instead, we shall start with continuum models of a classical ferromagnet. This approach allows one to obtain both linear and nonlinear excitations at the classical level and then quantize them. 

In Section \ref{sec:spin}, we discuss the dynamics of a single spin and introduce various mathematical tools. The same tools, suitably generalized, are used for the description of soliton dynamics in a one-dimensional ferromagnet in Section \ref{sec:ferromagnetic-wire}. Sec.~\ref{sec:Walker} deals with a more complex model of a domain wall of Schryer and Walker~\cite{Schryer1974}.

\section{Single magnetic dipole}
\label{sec:spin}

\subsection{Magnetic moment and angular momentum}

The building block of a macroscopic magnet is an atom with angular momentum $\mathbf J$ and magnetic dipole moment $\boldsymbol \mu$ related by the gyromagnetic ratio $\gamma$,
\begin{equation}
\boldsymbol \mu = \gamma \mathbf J.
\end{equation}
If the angular momentum comes solely from the electron spins, the gyromagnetic ratio is $\gamma = e /m_e c$, where $e<0$ is the electron charge and $m_e$ its mass. More generally, $\gamma = g e/2m_e c$, where the Lande $g$-factor is determined by the lengths of spin, orbital, and total angular momenta $S$, $L$, and $J$ \cite{Sakurai}. 

We shall treat both the angular momentum $\mathbf J$ and the magnetic moment $\boldsymbol \mu$ as classical physical variables of fixed lengths $J$ and $\mu$. Often both are expressed in terms of a unit vector $\mathbf m$ parallel to $\mathbf J$: 
\begin{equation}
\mathbf J = J \mathbf m,
\quad
\boldsymbol \mu = \gamma J \mathbf m, 
\quad
|\mathbf m| = 1.
\end{equation}
Because of the linear proportionality between the three quantities, we will casually refer to $\mathbf m$ as the magnetic moment or spin. 

\subsection{Spherical geometry}

\begin{figure}
\begin{center}
\includegraphics[width=0.4\columnwidth]{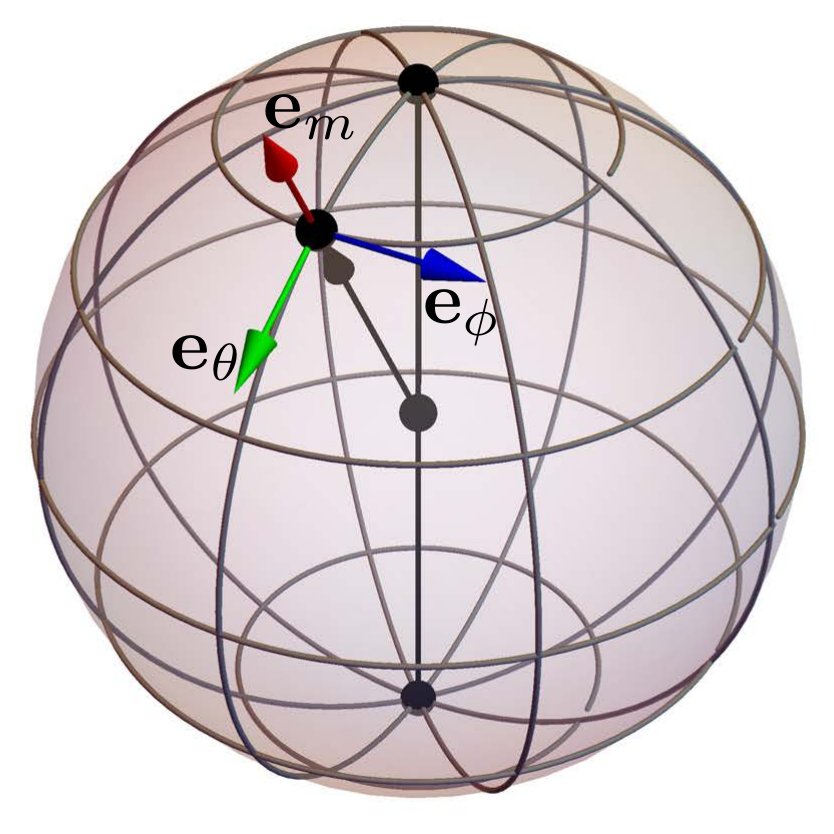}
\end{center}
\caption{The local frame defined by the unit vectors (\ref{eq:local-frame}).}
\label{fig:local-frame}
\end{figure}

To resolve the constraint of the unit length $|\mathbf m| = 1$, it is convenient to introduce two spherical angles:  the polar angle $\theta$ (co-latitude) and the azimuthal angle $\phi$ (longitude): 
\begin{equation}
\mathbf m = (\sin{\theta} \cos{\phi}, \sin{\theta} \sin{\phi}, \cos{\theta}).
\label{eq:theta-phi-def}
\end{equation}
They define a local frame with three mutually orthogonal unit vectors 
\begin{eqnarray}
&&\mathbf e_m = \mathbf m 
	= (\sin{\theta} \cos{\phi}, 
    	\sin{\theta} \sin{\phi}, 
        \cos{\theta}), 
\nonumber\\
&&\mathbf e_\theta = 
	\frac{\partial \mathbf m}{\partial \theta} 
	= (\cos{\theta} \cos{\phi}, 
    	\cos{\theta} \sin{\phi}, 
        -\sin{\theta}),
\label{eq:local-frame}
\\
&&\mathbf e_\phi 
	= \frac{1}{\sin{\theta}} 
    	\frac{\partial \mathbf m}{\partial \phi} 
	= (-\sin{\phi}, \cos{\phi}, 0),
\nonumber
\end{eqnarray}
pointing up, south, and east, respectively (Fig.~\ref{fig:local-frame}). 

Small deviations of magnetization from a given direction $\mathbf m$ can be written in the local frame as follows:
\begin{equation}
\delta \mathbf m 
	= \mathbf e_\theta \, \delta m_\theta 
    	+ \mathbf e_\phi \, \delta m_\phi,
\qquad
\delta m_\theta = \delta \theta, 
\quad
\delta m_\phi = \sin{\theta} \, \delta \phi.
\label{eq:m-theta-m-phi-def}
\end{equation}

\subsection{Precessional dynamics}

The dynamics of an atomic magnetic dipole is very different from that of a Newtonian particle. For a particle, an external force generates an acceleration proportional to the force and inversely proportional to its mass. In contrast, a magnetic dipole does not have inertia. Like a fast-spinning gyroscope, it precesses with a velocity proportional to the external torque. 

It is helpful to picture a magnetic dipole whose south pole is pivoted and the north pole is allowed to move on a sphere of radius $r$ so that its position is $\mathbf r = r \mathbf m$. The force from the potential energy may then be calculated as $\mathbf F = - \frac{d U}{d \mathbf r} = - \frac{1}{r}\frac{d U}{d \mathbf m}$ and its torque as $\boldsymbol \tau = \mathbf r \times \mathbf F = - \mathbf m \times \frac{d U}{d \mathbf m}$. The torque determines the rate of change of angular momentum $d \mathbf J / dt = J d \mathbf m / dt$, hence the equation of motion
\begin{equation}
J \dot{\mathbf m} = - \mathbf m \times \frac{d U}{d \mathbf m},
\label{eq:precession}
\end{equation}
where $\dot{\mathbf m}$ stands for $d \mathbf m / dt$. Note that the geometrical size of the dipole $r$ canceled out. 

It is worth keeping in mind that the derivative $d U/d \mathbf m$ is only taken in the directions locally tangential to the sphere $|\mathbf m| = 1$. In a formal sense, this derivative is defined as the coefficient in the Taylor expansion
\begin{equation}
U(\mathbf m + d \mathbf m) = U(\mathbf m) + \frac{d U}{d \mathbf m} \cdot d \mathbf m + \ldots,
\label{eq:dU/dm-defined}
\end{equation}
where $d \mathbf m$ is an infinitesimal displacement tangential to the sphere and thus transverse to $\mathbf m$. Even if $dU/d\mathbf m$ had a component parallel to $\mathbf m$, it would not contribute to the first order in the expansion (\ref{eq:dU/dm-defined}). For this reason, $dU/d\mmm$ should be understood as the transverse component of the energy derivative. 

\subsection{Example: Larmor precession} 

Consider the motion of a magnetic dipole in an external magnetic field $\mathbf h$. Its potential energy is $U = - \boldsymbol \mu \cdot \mathbf h = - \mu \mathbf m \cdot \mathbf h$. Upon substituting $U$ into the equation of motion (\ref{eq:precession}) we obtain 
\begin{equation}
J \dot{\mathbf m} = - \mu \mathbf h \times \mathbf m.
\end{equation}
The magnetic moment rotates uniformly about the direction of the field at the Larmor precession frequency 
\begin{equation}
\omega_L = \mu h/J = \gamma h.
\label{eq:Larmor-frequency}
\end{equation}
The angle between the magnetic moment and the field remains constant, implying conservation of energy $U = - \boldsymbol \mu \cdot \mathbf h$. 

To see this explicitly, rewrite the equation of motion in terms of spherical angles $\theta$ and $\phi$ (\ref{eq:theta-phi-def}): 
\begin{equation}
- J \sin{\theta} \, \dot{\phi} - \frac{\partial U}{\partial \theta} = 0,
\quad
J \dot{\theta} - \frac{1}{\sin{\theta}} \frac{\partial U}{\partial \phi} = 0.
\label{eq:eom-theta-phi-spin}
\end{equation}
For the field along the $z$ axis, $\mathbf h = (0,0, h)$, the potential energy $U = - \mu h \cos{\theta}$ is independent of the azimuthal angle $\phi$. Thus the polar angle $\theta$ remains unchanged during precession.

\subsection{Conservative and gyroscopic forces}

Conservation of energy is a general property of the precession equation (\ref{eq:precession}). After some simple algebra, it can be transformed into the form 
\begin{equation}
- J \dot{\mathbf m} \times \mathbf m - \frac{dU}{d\mathbf m} = 0.
\label{eq:2nd-law-spin}
\end{equation}
Returning to the picture of a dipole pivoted at the south pole, we may interpret this equation of motion as Newton's second law for a massless particle confined to move on a unit sphere (hence no inertial term proportional to $\ddot{\mathbf m}$). The second in Eq.~(\ref{eq:2nd-law-spin})  obviously represents a conservative force. 

What does the first term represent? It is similar to the Lorentz force acting on a moving electric charge in a magnetic field as it is proportional to the velocity $\dot{\mathbf m}$ and is orthogonal to it. We thus may interpret the first term in Eq.~(\ref{eq:2nd-law-spin}) as a Lorentz force acting on a particle with unit electric charge moving on the unit sphere $|\mathbf m| = 1$ in a radial ``magnetic field'' $\mathbf b(\mathbf m) = - J \mathbf m$, as if there were a magnetic monopole of strength $J$ at the sphere's center. (We use the scare quotes to distinguish the fictitious ``magnetic field'' $\mathbf b$ on the unit sphere $|\mathbf m|=1$ from the physical magnetic field $\mathbf h$.)

The ``Lorentz force'' in Eq.~(\ref{eq:2nd-law-spin}) does no work and thus cannot be described in terms of potential energy. It belongs to a class of \emph{gyroscopic} forces first discussed by Kelvin \cite{Thomson:1879}. Precession of a spin is indeed reminiscent of the motion of a fast-spinning gyroscope. 

Eq.~(\ref{eq:2nd-law-spin}) can be used to establish conservation of potential energy during motion: 
\begin{equation}
\frac{dU}{dt} 
	= \dot{\mathbf m} \cdot \frac{dU}{d\mathbf m} 
	= \dot{\mathbf m} \cdot (- J \dot{\mathbf m} \times \mathbf m) = 0.
\end{equation}

\subsection{Lagrangian}

The equation of motion (\ref{eq:2nd-law-spin}) can be obtained by minimization of action $S = \int L \, dt$. The Lagrangian 
\begin{equation}
L = \mathbf a(\mathbf m) \cdot \dot{\mathbf m} - U(\mathbf m)
\label{eq:L-spin}
\end{equation}
has no kinetic energy~\cite{Altland-Simons}. The first term gives rise to the gyroscopic (``Lorentz'') force and contains a gauge potential $\mathbf a(\mathbf m)$ whose curl gives the ``magnetic field'' of the monopole, 
\begin{equation}
\nabla_{\mathbf m} \times \mathbf a(\mathbf m) = \mathbf b(\mathbf m) = - J \mathbf m.
\end{equation}

The first term in the Lagrangian (\ref{eq:L-spin}) is peculiar for a number of reasons. 

\begin{enumerate}

\item Its action 
\begin{equation}
S_g = \int \mathbf a(\mathbf m) \cdot \dot{\mathbf m} \, dt 
	= \int \mathbf a(\mathbf m) \cdot d \mathbf m
\end{equation}
is independent of how fast the vector $\mathbf m$ moves on the unit sphere and depends just on the geometry of its path. Hence the name \emph{geometric action} for $S_g$ and geometric (or Berry) phase for $S_g/\hbar$ \cite{Shapere:1989}. 

\item This term is not uniquely defined. A gauge transformation 
\begin{equation}
\mathbf a(\mathbf m) \mapsto \mathbf a(\mathbf m) + \nabla_{\mathbf m} \, \chi(\mathbf m),
\end{equation}
where $\chi(\mathbf m)$ is an arbitrary scalar function on the unit sphere, changes the vector potential $\mathbf a(\mathbf m)$ but leaves the ``magnetic field'' $\mathbf b(\mathbf m)$ the same. 

\item Strictly speaking, the magnetic field of a monopole $\mathbf b(\mathbf m)$ cannot be represented by a gauge potential because this field configuration is not divergence-free: it has a source with magnetic charge $-J$ at the origin $\mathbf m=0$ \cite{Altland-Simons, Haldane1986}. Although the magnetic monopole is not accessible, we can still observe that there is a net ``magnetic flux'' $-4\pi J$ by integrating the ``field'' over the unit sphere. 

\item As a result of this problem, any choice of the gauge potential $\mathbf a(\mathbf m)$ contains a singularity. For example, the gauge potential \cite{Haldane1986} 
\begin{equation}
\mathbf a(\mathbf m)  
	= J \, \frac{\mathbf m_s \times \mathbf m}{1 - \mathbf m_s \cdot \mathbf m},
\label{eq:a-string}
\end{equation}
has a singularity at $\mmm = \mmm_s$. It is associated with a localized ``magnetic flux'' of strength $+4\pi J$, which compensates the uniform flux $-4\pi J$ spread over the sphere. If we were allowed to explore the entire 3-dimensional space of $m$, we would discover that there is a Dirac string carrying ``magnetic flux'' $+4\pi J$ from the origin to infinity in the direction $\mathbf m_s$. With the Dirac string attached, the ``magnetic field'' becomes solenoidal, which makes it possible to describe it in terms of a vector potential. 

\item Vector potential (\ref{eq:a-string}) describes the ``magnetic field'' of a monopole $\mathbf b(\mathbf m) = - J \mathbf m$ \emph{almost} everywhere on the unit sphere $\mathbf m$, with the exception of point $\mathbf m_s$, where the ``magnetic flux'' of the Dirac string pierces the sphere. As long as the trajectory $\mathbf m(t)$ stays away from the Dirac string, we can use the gauge potential. 

\end{enumerate}

To complete this section, we write down the gauge term in spherical angles for the azimuthally symmetric gauge choices with the Dirac string attached to the north and south poles, $\mmm_s = (0,0,\pm1)$: 
\begin{equation}
\mathbf a(\mathbf m) \cdot \dot{\mathbf m} = J (\cos{\theta} \pm 1) \dot{\phi}.
\label{eq:gauge-standard}
\end{equation}
These are the two most commonly used gauge choices \cite{Altland-Simons}. 

\subsection{Example: precession near a potential minimum}
\label{sec:spin-precession-near-minimum}

Choose the axes so that the energy has a minimum at the north pole, $\mathbf m_0 = (0,0,1)$. In the vicinity of the energy minimum, 
\begin{equation}
\mmm = (m_x, \, m_y, \, \sqrt{1-m_x^2-m_y^2}), 
\end{equation}
we may expand the potential energy in powers of the transverse deviations $m_x \ll 1$ and $m_y \ll 1$. To the second order,  
\begin{equation}
U(\mathbf m) = U(\mathbf m_0) 
	+ \frac{K_x m_x^2}{2} + \frac{K_y m_y^2}{2} 
    + \ldots
\end{equation}
Here $K_x$ and $K_y$ are positive constants and the omitted terms are of higher orders in the transverse components of magnetization. Choose the gauge potential (\ref{eq:a-string}) with the singularity at the south pole, $\mathbf m_s = (0,0,-1)$, so that $\mathbf a \approx (J m_y/2, -J m_x/2, 0) + \ldots$ The resulting Lagrangian 
\begin{equation}
L \approx
	\left[
    	J (\dot{m}_x m_y - \dot{m}_y m_x) 
        - K_x m_x^2 - K_y m_y^2
    \right]/2
    + \ldots
\end{equation}
yields the equations of motion for $m_x$ and $m_y$: 
\begin{equation}
- J \dot{m}_y - K_x m_x = 0, 
\quad
J \dot{m}_x - K_y m_y = 0.
\end{equation}
The spin follows an elliptical trajectory $K_x m_x^2 + K_y m_y^2 = \mathrm{const}$ precessing at the frequency $\omega = - \sqrt{K_x K_y}/J$.

\subsection{Dissipative force}

In the real world, a magnetic dipole will not precess about the magnetic field forever and will eventually line up with it, settling into a minimum of potential energy. Energy dissipation can be caused by various processes, e.g., radiation of electromagnetic waves or friction. We shall add a simple phenomenological term to the equations of motion that represents---in the analogy with a particle on a sphere---a dissipative force $- \alpha J \dot{\mathbf m}$ proportional to the velocity $\dot{\mathbf m}$ and directed against it. Eq.~(\ref{eq:2nd-law-spin}) now reads 
\begin{equation}
- J \dot{\mathbf m} \times \mathbf m - \frac{dU}{d\mathbf m} - \alpha J \dot{\mathbf m} = 0.
\label{eq:2nd-law-spin-with-friction}
\end{equation}
It expresses the balance of gyroscopic, conservative, and dissipative forces acting on the dipole. The dimensionless constant $\alpha > 0$ introduced by Gilbert \cite{Gilbert2004} characterizes the strength of energy dissipation. It is usually small, $\alpha \ll 1$, with typical values ranging from $10^{-4}$ in insulating magnets to $10^{-2}$ in metallic ones. 

The equation of motion, expressed in spherical angles, reads
\begin{eqnarray}
- J \sin{\theta} \, \dot{\phi} 
	- \frac{\partial U}{\partial \theta} 
    - \alpha J \dot{\theta} 
    = 0,
\nonumber\\
J \dot{\theta} 
	- \frac{1}{\sin{\theta}} \frac{\partial U}{\partial \phi} 
    - \alpha J \sin{\theta} \, \dot{\phi}
    = 0.
\label{eq:2nd-law-spin-with-friction-spherical-angles}
\end{eqnarray}

In the presence of the dissipative force, potential energy is dissipated at the rate
\begin{equation}
\frac{dU}{dt} 
	= \dot{\mathbf m} \cdot \frac{dU}{d\mathbf m} 
	= \dot{\mathbf m} \cdot (- J \dot{\mathbf m} \times \mathbf m - \alpha J \dot{\mathbf m}) = - \alpha J |\dot{\mathbf m}|^2 \leq 0.
\end{equation}

\subsection{Example: Larmor precession with dissipation}

\begin{figure}
    \centering
   \includegraphics[width=0.4\columnwidth]{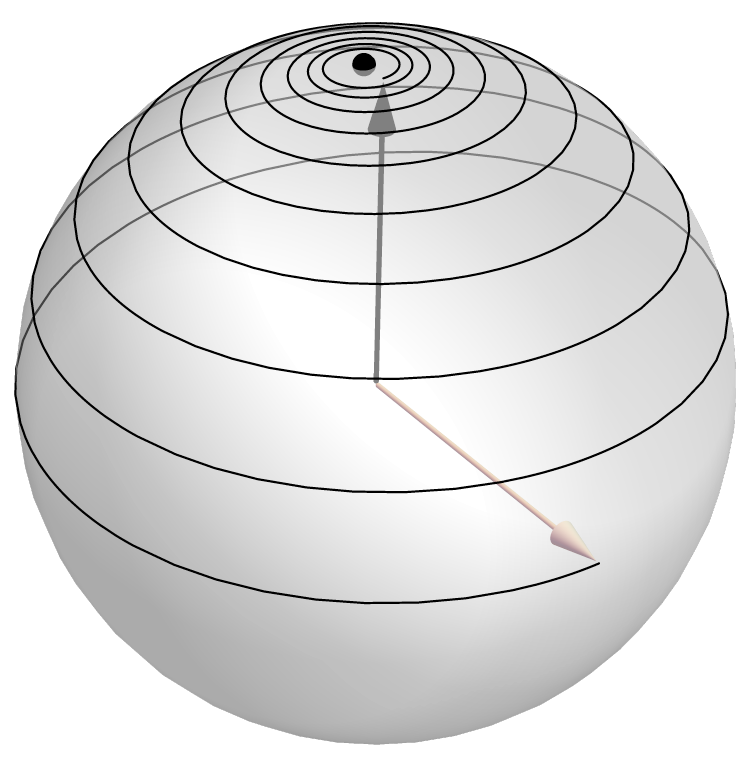}
    \caption{Larmor precession of a spin in an external magnetic field in the presence of dissipation. The energy minimum is at the north pole (black point). The spin trajectory is shown for $1/(2\alpha)$ Larmor periods with the starting point on the equator. The Gilbert damping $\alpha=0.05$; the gyromagnetic ratio $\gamma>0$.}
    \label{fig:spin-in-external-field}
\end{figure}

We return to the problem of a dipole in a magnetic field, this time with dissipation. With energy $U = - \mu h \cos{\theta}$, Eq.~(\ref{eq:2nd-law-spin-with-friction-spherical-angles}) reads 
\begin{eqnarray}
-J \sin{\theta} \, \dot{\phi} - \mu h \sin{\theta} - \alpha J \dot{\theta} = 0,
\nonumber\\
J \dot{\theta} - \alpha J \sin{\theta} \, \dot{\phi} = 0.
\end{eqnarray}
The introduction of dissipation leads to a slight slowdown of the Larmor precession and to a gradual approach to the energy minimum: 
\begin{eqnarray}
\phi(t) = \phi(0) - \tilde{\omega}_L t, 
\quad 
\tilde{\omega}_L = \frac{\omega_L}{1+\alpha^2},
\nonumber\\
\cos{\theta(t)} = \tanh{\left(\alpha \tilde{\omega}_L (t-t_0)\right)}.
\label{eq:precession-spin-with-friction}
\end{eqnarray}
Here $\omega_L = \gamma h$ is the bare Larmor frequency (\ref{eq:Larmor-frequency}) and $t_0$ is the time when the spin lies in the equatorial plane, $\cos{\theta}=0$. We see that the azimuthal motion is precessional and fast, with the frequency of order $\alpha^0$, whereas the polar motion is relaxational and slow, with the relaxation rate of order $\alpha$. Fig.~\ref{fig:spin-in-external-field} shows the trajectory of a spin for $1/(2\alpha)$ Larmor periods starting at the equator and approaching the energy minimum at the north pole.  

\subsection{Transformation to canonical variables}
\label{sec:spin-Hamilton}

Although our physical quantities---magnetization and angular momentum---are represented by a three-dimensional vector $\mathbf m$, its three components are not independent because of the fixed length $|\mathbf m| = 1$. It may sometimes be convenient to work with two independent variables, which can be introduced in a number of ways. 

One possible route is to use a pair of canonical variables $(q,p)$, one of which is a coordinate and the other its conjugate momentum. To obtain them, we may use the Lagrangian in one of the standard gauges (\ref{eq:gauge-standard}),
\begin{equation}
L = J (\cos{\theta} \pm 1) \dot{\phi} - U(\theta, \phi). 
\label{eq:L-spin-gauge-standard}
\end{equation}
We choose the azimuthal angle $\phi$ as the coordinate and obtain the corresponding canonical momentum in the usual way: 
\begin{equation}
q = \phi,
\quad
p = \frac{\partial L}{\partial \dot{\phi}} = J (\cos{\theta} \pm 1) = J_z \pm J. 
\end{equation}
As one might expect, the momentum conjugate to $\phi$ is (up to an additive constant) the angular momentum $J_z = J \cos{\theta}$. It is somewhat alarming that the polar angle $\theta$ does not have a conjugate momentum (no $\dot{\theta}$ term in the Lagrangian). We will address this problem in a later section. For now, we shall dismiss this concern by noting that $\theta$ is present in the canonical momentum so that both angles are represented in the canonical pair $(q, p)$. 

We now realize that our two-dimensional space is a \emph{phase space}. The number of states in a phase space is proportional to phase-space volume:
\begin{equation}
d\Gamma = \frac{dq \, dp}{2\pi \hbar} 
	= \frac{J}{2\pi \hbar} \sin{\theta} \, d\theta \, d\phi.
\end{equation}
Integrating over the sphere, we obtain the total number of states
\begin{equation}
\Gamma 
	= \frac{J}{2\pi \hbar} \int_0^\pi \sin{\theta} \, d\theta 
    	\int_0^{2\pi} d\phi 
    = \frac{2J}{\hbar} = 2j.
\end{equation}
where $j = J/\hbar$ is the length of angular momentum in the units of $\hbar$. The exact answer, which we have learned in a quantum mechanics course, is $\Gamma = 2j+1$. Our approximate answer, obtained within the scope of classical mechanics, is correct in the limit of a large length of angular momentum, $j \gg 1$. 

With canonical variables identified, we have the Poisson bracket. In particular, 
\begin{equation}
\{\theta,\phi\} = - \{\phi,\theta\}
	= \frac{\partial \theta}{\partial q} \frac{\partial \phi}{\partial p}
    	- \frac{\partial \theta}{\partial p} \frac{\partial \phi}{\partial q}
    = \frac{1}{J\sin{\theta}}.
\end{equation}

The Hamiltonian is obtained in a standard way, 
\begin{equation}
H = p \dot{q} - L = U(\theta,\phi).
\end{equation}
It contains potential energy only. The $\dot{\phi}$ term in the Lagrangian (\ref{eq:L-spin-gauge-standard}) is \emph{not} kinetic energy. Hamilton's equations of motion, 
\begin{equation}
\dot{q} = \{q,H\} = \partial H/\partial p, 
\quad
\dot{p} = \{p,H\} = - \partial H/\partial q,
\label{eq:Hamilton}
\end{equation}
reproduce the equations of motion for the spherical angles (\ref{eq:eom-theta-phi-spin}).

\subsection{Transformation to general variables}

We may not always want to use pairs of canonically conjugate variables and instead perform a more general transformation from $\mathbf m$ to a pair of coordinates $(q^1, q^2)$, e.g., $q^1 = \theta$ and $q^2 = \phi$. Equations of motion for the new variables can be obtained by starting from the equation (\ref{eq:2nd-law-spin-with-friction}) expressing force balance. 

To obtain the equations of motion for the new variables $\{q^i\}$, where $i = 1,2$, we first note that the time evolution of $\mathbf m(q^1,q^2)$ comes solely through the time evolution of the coordinates: $\dot{\mathbf m} = \dot{q}^j \, \partial \mathbf m/\partial q^j$, where summation is implied over the doubly repeated index $j$. We substitute this expression into Eq.~(\ref{eq:2nd-law-spin-with-friction}) and multiply the result by $\partial \mathbf m/\partial q^i$ to obtain an equation of motion for the new coordinates: 
\begin{equation}
G_{ij} \dot{q}^j - \partial U/\partial q^i - D_{ij} \dot{q}^j = 0.
\label{eq:2nd-law-spin-with-friction-general-variables}
\end{equation}
It expresses the balance of the gyroscopic, conservative, and dissipative forces for each coordinate $q^i$. The coefficients $G_{ij}$ and $D_{ij}$ of the antisymmetric gyroscopic and symmetric dissipative tensors are defined as follows: 
\begin{equation}
G_{ij} = - J \mathbf m \cdot 
	\left( 
    	\frac{\partial \mathbf m}{\partial q^i} 
        	\times \frac{\partial \mathbf m}{\partial q^j} 
    \right), 
\quad
D_{ij} = \alpha J \, 
	\frac{\partial \mathbf m}{\partial q^i}
    	\cdot \frac{\partial \mathbf m}{\partial q^j}.
\end{equation}
The gyroscopic tensor $G_{ij}$ is the inverse of the of Poisson tensor $\Pi^{ij} = \{q^i,q^j\}$, i.e., $G_{ij} \Pi^{jk} = \delta_i^k$ \cite{Gonzalez2022}. It can be obtained as the curl of a gauge potential $A_i$: 
\begin{equation}
G_{ij} 
= \frac{\partial A_j}{\partial q^i} 
- \frac{\partial A_i}{\partial q^j},
\quad
A_i = \mathbf a(\mmm) 
    \cdot \frac{\partial \mmm}{\partial q^i}.
\label{eq:G-via-A-spin}
\end{equation}
The Lagrangian, expressed in terms of the new coordinates, reads
\begin{equation}
L = A_i(q) \dot{q}^i - U(q).    
\end{equation}
The first term of the Lagrangian gives rise to the geometrical action 
\begin{equation}
S_g = \int A_i \dot{q}^i dt = \int A_i \, dq^i.    
\end{equation}

The number of states in an infinitesimal rectangle with sides $(dq^1, dq^2)$ is 
\begin{equation}
d\Gamma 
	= \frac{1}{2\pi \hbar} |G_{12}| \, dq^1 \, dq^2 
    =  \sqrt{\det G} \,
    	\prod_{i=1}^2 \frac{dq^i}{\sqrt{2\pi \hbar}}.
\label{eq:dGamma-spin}
\end{equation}

For the choice of coordinates $q^1 = \theta$ and $q^2 = \phi$, the nonzero components of the tensors are
\begin{equation}
G_{\theta\phi} = - G_{\phi\theta} = - J \sin{\theta}, 
\quad
D_{\theta\theta} = \alpha J, 
\quad 
D_{\phi\phi} = \alpha J \sin^2{\theta}. 
\end{equation}
Upon substituting these tensor components into Eq.~(\ref{eq:2nd-law-spin-with-friction-general-variables}), we obtain the equations of motion for the spherical angles (\ref{eq:2nd-law-spin-with-friction-spherical-angles}).

\subsection{Rayleigh's dissipation function}

Dissipative forces cannot be described as a term in a Lagrangian. It is an emergent force, arising from interaction of a macroscopic object with numerous microscopic degrees of freedom. In thermodynamics, it plays a major role in the relaxation of the physical system toward thermal equilibrium. In that context, the dissipative force can be obtained from the Rayleigh dissipation function $R$ quadratic in velocities \cite{LL-I},
\begin{equation}
R = \frac{1}{2} \alpha J |\dot{\mathbf m}|^2
	= \frac{1}{2} D_{ij} \dot{q}^i \dot{q}^j.
\label{eq-R-spin}
\end{equation}
The modified Euler-Lagrange equations are
\begin{equation}
\frac{\partial L}{\partial q^i} 
- \frac{d}{dt} \frac{\partial L}{\partial \dot{q}^i}
- \frac{\partial R}{\partial \dot{q}^i} = 0.
\end{equation}
The last term is the dissipative force.

\subsection{Example: strong easy-plane anisotropy}
\label{sec:spin-Doring}

Let us discuss the dynamics of a spin whose potential energy 
\begin{equation}
U(\theta, \phi) = -\frac{K_\theta}{2} \sin^2{\theta} + V(\theta,\phi)
\end{equation}
is dominated by the easy-plane anisotropy term with $K_\theta>0$. Whatever the starting state, after a period of initial relaxation the spin will find itself near the equatorial circle $\theta = \pi/2$. Its long-term dynamics will be influenced by the azimuthal potential landscape defined by $V(\pi/2,\phi)$. 

To get a first look at the long-term dynamics, consider the motion in the vicinity of the potential energy minimum at $\theta = \pi/2$ and some $\phi = \phi_0$, where the potential energy is approximately quadratic in the deviations $\delta \theta$ and $\delta \phi$ from the minimum: 
\begin{equation}
U(\pi/2 + \delta \theta, \phi_0 + \delta \phi) 
	\approx \frac{K_\theta \delta \theta^2}{2} + \frac{K_\phi \delta \phi^2}{2},
\end{equation}
where $K_\phi \ll K_\theta$. Equations of motion (\ref{eq:2nd-law-spin-with-friction-spherical-angles}), linearized in the deviations from equilibrium, read 
\begin{equation}
- J \, \delta \dot{\phi} - K_\theta \, \delta \theta - \alpha J \, \delta \dot{\theta} = 0, 
\quad
J \, \delta \dot{\theta} - \, K_\phi \delta \phi - \, \alpha J \delta \dot{\phi} = 0.
\end{equation}

In the absence of dissipation, $\alpha = 0$, we obtain elliptical precession near the equatorial plane with the frequency $\omega = \sqrt{K_\theta K_\phi}/J$ and the amplitude ratio $\delta\theta_0/\delta\phi_0 = \sqrt{K_\phi/K_\theta} \ll 1$. Turning on dissipation at first makes the precession underdamped, with the relaxation rate $\Gamma \approx \alpha K_\theta/2J$ in the limit $\alpha \ll 2\sqrt{K_\phi/K_\theta}$. When the damping constant exceeds the critical value $\alpha_c = 2\sqrt{K_\theta K_\phi}/(K_\theta - K_\phi) \ll 1$, the spin dynamics becomes overdamped, with two distinct relaxation rates, $\Gamma_1 \approx \alpha K_\theta/J$ and $\Gamma_2 \approx K_\phi/\alpha J \ll \Gamma_1$ in the limit $2\sqrt{K_\phi/K_\theta} \ll \alpha \ll 1$.   

Our simple analysis of the final approach to equilibrium demonstrates that, even when dissipation is nominally weak, $\alpha \ll 1$, spin dynamics does not necessarily retain its precessional character and can be purely relaxational. An apparently small damping constant $\alpha \ll 1$ may represent relatively strong dissipation when it is compared to another small parameter, the ratio of azimuthal and polar anisotropies $\sqrt{K_\phi/K_\theta}$. Put another way, increasing the strength of the easy-plane anisotropy $K_\theta$ at a fixed $\alpha \ll 1$ will eventually bring us into a regime where the spin motion separates into two independent components, fast relaxation, during which the spin approaches the equatorial plane, and slow relaxation within the equatorial plane. (The fast mode include azimuthal as well as polar motion.)

Next we shall derive the long-term dynamics of spin motion in a more general setting, when the polar angle is already near equilibrium, $\theta = \pi/2 + \delta \theta$ with a small $\delta \phi$, but $\phi$ may not be. The Lagrangian can then be approximated as follows:
\begin{equation}
L \approx - J \, \delta \theta \, \dot{\phi} - K_\theta \delta \theta^2/2 - V(\pi/2,\phi).
\label{eq:L-spin-easy-plane}
\end{equation}
In the Rayleigh dissipation function, 
\begin{equation}
R \approx \alpha J(\delta \dot{\theta}^2 + \dot{\phi}^2)/2 
	\approx \alpha J \dot{\phi}^2/2,
\label{eq:R-spin-easy-plane}
\end{equation}
we have neglected the contribution of polar motion to damping because it is restricted by the strong easy-plane anisotropy, so $\delta \dot{\theta} \ll \dot \phi$. This is an important general point, so let us emphasize it: \emph{dissipation is dominated by soft modes; dissipation from hard modes can often be neglected}. 

From the Lagrangian (\ref{eq:L-spin-easy-plane}) and Rayleigh function (\ref{eq:R-spin-easy-plane}) we obtain the equations of motion: 
\begin{equation}
- J \dot{\phi} - K_\theta \delta \theta = 0, 
\quad
J \delta \dot{\theta} - \partial V/\partial \phi - \alpha J \dot{\phi} = 0.
\end{equation}
The first equation lets us express the polar angle (a hard mode of no interest to us) in terms of the azimuthal angle (the soft mode we are interested in). Then the second equation will be for the azimuthal angle alone: 
\begin{equation}
I \ddot{\phi} - \alpha J \dot{\phi} - \partial V/\partial \phi = 0,
\label{eq:eom-spin-easy-plane-phi}
\end{equation}
where we introduced a ``moment of inertia'' $I = J^2/K_\theta$. 

By integrating out the hard mode $\delta \theta$, we have generated inertia (mass) for the soft mode $\phi$. This is a sufficiently frequent occurrence in ferromagnetic dynamics to merit a proper name, the D{\"o}ring mass \cite{Doring1948}. The Lagrangian of the azimuthal angle alone (with the polar angle integrated out) acquires kinetic energy: 
\begin{equation}
L(\phi) = I \dot{\phi}^2/2 - V(\pi/2,\phi).
\end{equation}
This elegant result hinges on neglect of dissipation for the hard mode $\delta \theta$. Had we not neglected it, the equation relating $\theta$ to $\phi$ would be more involved---non-local in time---and this simple picture would not have emerged. The non-locality in time is negligible on time scales longer than $\tau_1 = 1/\Gamma_1$.

Returning to the equation of motion for the azimuthal angle (\ref{eq:eom-spin-easy-plane-phi}), we see that it describes a massive rotator subject to an external potential and dissipative friction. On long time scales, we may neglect inertia and obtain a simpler equation of dissipative dynamics, 
\begin{equation}
- \alpha J \dot{\phi} - \partial V/\partial \phi = 0.
\label{eq:eom-spin-easy-plane-phi-one-mode}
\end{equation}
The azimuthal velocity $\dot{\phi}$ is proportional to the potential torque $- \partial V/\partial \phi$. Doing so is permissible on time scales longer than the characteristic acceleration time $\tau_1 = I/\alpha J = J/\alpha K_\theta = 1/\Gamma_1$. 

We should check that our long-term dynamics reproduces the previously obtained approach to equilibrium, when $\phi = \phi_0 + \delta \phi$. For small $\delta \phi$, the equatorial torque $- \partial V/\partial \phi \approx - k \, \delta \phi$ and we recover exponential relaxation of $\delta \phi$ at the rate $\Gamma_2 = K_\phi/\alpha J$, in agreement with our preliminary analysis. 

On a final note, we observe that the long-term dynamics of the soft mode $\phi$ given by Eq.~(\ref{eq:eom-spin-easy-plane-phi-one-mode}) can be obtained directly from the Lagrangian (\ref{eq:L-spin-easy-plane}) and Rayleigh dissipation function (\ref{eq:R-spin-easy-plane}) by merely ignoring the hard mode $\delta\theta$. This is a general principle that we will apply in many contexts in what follows: \emph{the response of a system to weak external perturbations is dominated by its soft modes}.

\subsection{Recap: equations of motion}

We end the discussion of a single magnetic dipole by gathering in one place different forms of its equations of motion, all of which are equivalent to one another. 

\begin{itemize}

\item 
This form of the equation expresses the balance of the gyroscopic, conservative, and dissipative forces acting on a magnetic dipole: 
\begin{equation}
- J \dot{\mathbf m} \times \mathbf m - \frac{dU}{d\mathbf m} - \alpha J \dot{\mathbf m} = 0.
\label{eq:2nd-law-spin-with-friction-recap}
\end{equation}

\item This equation expresses the rate of change of angular momentum in terms of conservative and dissipative torques acting on the dipole: 
\begin{equation}
J \dot{\mathbf m} = - \mathbf m \times \frac{dU}{d \mathbf m} - \alpha J \mathbf m \times \dot{\mathbf m}.
\label{eq:precession-with-friction-recap}
\end{equation}

\item The previous equation is often written in terms of an ``effective field'' $\mathbf h_\mathrm{eff} = - dU/d\boldsymbol \mu$ as follows:
\begin{equation}
\dot{\mathbf m} = - \gamma \mathbf h_\mathrm{eff} \times \mathbf m 
	- \alpha \mathbf m \times \dot{\mathbf m}, 
\quad
\mathbf h_\mathrm{eff} = - \frac{1}{\mu} \frac{dU}{d\mathbf m}.
\label{eq:LLG-spin}
\end{equation}
This form of the equation of motion is known as the Landau-Lifshitz-Gilbert (LLG) equation. 

\item The LLG equation (\ref{eq:LLG-spin}) contains two terms with a time derivative $\dot{\mathbf m}$, which is inconvenient when the equation is integrated numerically. Through simple algebra, it can be converted into the form used originally by Landau and Lifshitz, 
\begin{equation}
(1+\alpha^2) \dot{\mathbf m} 
	= - \gamma \mathbf h_\mathrm{eff} \times \mathbf m 
      - \alpha |\gamma| \mathbf h_\mathrm{eff}. 
\label{eq:LL-spin}
\end{equation}
This form of the equation contains a single term with the time derivative $\dot{\mathbf m}$ and is used in numerical micromagnetic solvers such as OOMMF and MuMax. One cautionary remark is on order. The effective field $\mathbf h_\mathrm{eff}$ defined in Eq.~(\ref{eq:LLG-spin}) is understood as a derivative $dU(\mathbf m)/d\mathbf m$ with respect to \emph{transverse} variations of $\mathbf m$, see Eq.~(\ref{eq:dU/dm-defined}). A sloppy evaluation may produce a longitudinal component of $\mathbf h_\mathrm{eff}$ parallel to $\mathbf m$ and create a problem for the Landau-Lifshitz equation (\ref{eq:LL-spin}), changing the length of the unit vector $\mathbf m$. Just to be on the safe side, one may replace $\mathbf h_\mathrm{eff}$ in the second term of Eq.~(\ref{eq:LL-spin}) with its transverse part $\mathbf m \times (\mathbf h_\mathrm{eff} \times \mathbf m)$. The same applies to Eq.~(\ref{eq:2nd-law-spin-with-friction-recap}). 

\item 
Written in terms of spherical angles, the equations of motion are
\begin{eqnarray}
- J \sin{\theta} \, \dot{\phi} 
	- \frac{\partial U}{\partial \theta} 
    - \alpha J \dot{\theta} 
    = 0,
\nonumber\\
J \dot{\theta} 
	- \frac{1}{\sin{\theta}} \frac{\partial U}{\partial \phi} 
    - \alpha J \sin{\theta} \, \dot{\phi}
    = 0.
\label{eq:2nd-law-spin-with-friction-spherical-angles-recap}
\end{eqnarray}

\item 
For arbitrary coordinates $q^1$ and $q^2$ parametrizing the unit vector $\mathbf m$, the equations of motion for these coordinates again expresses the balance of the gyroscopic, conservative, and dissipative forces associated with each coordinate: 
\begin{equation}
G_{ij} \dot{q}^j - \partial U/\partial q^i - D_{ij} \dot{q}^j = 0.
\label{eq:2nd-law-spin-with-friction-general-variables-recap}
\end{equation}
The antisymmetric gyroscopic and symmetric dissipative tensors are 
\begin{equation}
G_{ij} = - J \mathbf m \cdot 
	\left( 
    	\frac{\partial \mathbf m}{\partial q^i} 
        	\times \frac{\partial \mathbf m}{\partial q^j} 
    \right), 
\quad
D_{ij} = \alpha J \, 
	\frac{\partial \mathbf m}{\partial q^i}
    	\cdot \frac{\partial \mathbf m}{\partial q^j}.
\label{eq:G-and-D-spin-recap}
\end{equation}

\end{itemize}

\section{Domain wall in a ferromagnetic wire}
\label{sec:ferromagnetic-wire}

\subsection{Heisenberg spin chain}

Theory of a single magnetic dipole can be readily extended to that of many dipoles. A convenient way to accomplish that is to start with the Lagrangian of a single spin (\ref{eq:L-spin}) and to generalize it to many spins $\{\mathbf m_n\}$: 
\begin{equation}
L = \sum_n \mathbf a(\mathbf m_n) \cdot \dot{\mathbf m}_n - U(\{\mathbf m_n\}), 
\label{eq:L-N-spins}
\end{equation}
where once again the vector potential $\mathbf a(\mathbf m)$ describes the field of a magnetic monopole, $\nabla_\mathbf m \times \mathbf a(\mathbf m) = - J \mathbf m$, and $U(\{\mathbf m_n\})$ is the potential energy of the system. The Rayleigh function is generalized in the same way, 
\begin{equation}
R = \frac{1}{2} \sum_n \alpha J |\dot{\mathbf m}_n|^2.
\end{equation}
Then the Euler-Lagrange equations of motion for every spin $\mathbf n$, 
\begin{equation}
\frac{\partial L}{\partial \mathbf m_n} 
	- \frac{d}{dt} \frac{\partial L}{\partial \dot{\mathbf m}_n}
    - \frac{\partial R}{\partial \dot{\mathbf m}_n}
    = 0,
\end{equation}
yields an obvious generalization of single-spin dynamics (\ref{eq:precession-with-friction-recap}), 
\begin{equation}
J \dot{\mathbf m}_n 
= - \mathbf m_n \times \frac{\partial U}{\partial \mathbf m_n} 
- \alpha J \mathbf m_n \times \dot{\mathbf m}_n,
\label{eq:precession-N-spins-with-friction}
\end{equation}
or its variants, Eqs.~(\ref{eq:2nd-law-spin-with-friction-recap}) and (\ref{eq:LLG-spin}). The effective field acting on spin $\mathbf m_n$ is $\mathbf h^{\mathrm{eff}}_n = - \mu^{-1} \, \partial U/\partial \mathbf m_n$. 

Here we examine a Heisenberg chain with magnetic dipoles arranged periodically with lattice constant $a$. Adjacent dipoles $\mathbf m_n$ and $\mathbf m_{n+1}$ interact with energy $- A \, \mathbf m_n \cdot \mathbf m_{n+1}$, where $A$ is the exchange constant: 
\begin{equation}
U = - A \sum_n \mathbf m_n \cdot \mathbf m_{n+1}. 
\label{eq:U-Heisenberg-chain}
\end{equation}
The equation of motion for dipole $\mathbf m_n$ is 
\begin{equation}
J \dot{\mathbf m}_n = 
 A \mathbf m_n \times (\mathbf m_{n-1} + \mathbf m_{n+1})
- \alpha J \mathbf m_n \times \dot{\mathbf m}_n.
\label{eq:dynamics-Heisenberg-chain}
\end{equation}

An equilibrium state, $\dot{\mathbf m}_n = 0$, is reached when all dipoles line up along the same direction, $\mathbf m_n = \mathbf m$. Such a state clearly minimizes the potential energy (\ref{eq:U-Heisenberg-chain}) for $A>0$. The common direction of magnetization $\mathbf m$ is arbitrary. 

It is instructive to investigate the dynamics of magnetization close to equilibrium, $\mathbf m_n = \mathbf m + \delta \mathbf m_n$, where infinitesimal deviations from equilibrium $\delta \mathbf m_n$ are transverse to $\mathbf m$ in order to preserve the unit length of $\mathbf m_n$. 

It is convenient to introduce a reference frame with three unit vectors $\mathbf e_x$, $\mathbf e_y$, and $\mathbf e_z$ forming a right triple, $\mathbf e_i \times \mathbf e_j = \epsilon_{ijk} \mathbf e_k$. We align one of them with the equilibrium direction, $\mathbf e_z = \mathbf m$. Then $\delta \mathbf m_n = m_{n,x} \, \mathbf e_x + \delta m_{n,y} \, \mathbf e_y$. After expanding Eq.~(\ref{eq:dynamics-Heisenberg-chain}) to the first order in $\delta \mathbf m_n$, we obtain the dynamical equation for the deviations: 
\begin{eqnarray}
J \dot{m}_{n,x} &=& 
- A(m_{n-1,y} - 2 m_{n,y} + m_{n+1,y})
+ \alpha J \dot{m}_{n,y},
\nonumber\\
- J \dot{m}_{n,y} &=& 
- A(m_{n-1,x} - 2 m_{n,x} + m_{n+1,x})
+ \alpha J \dot{m}_{n,x}.
\end{eqnarray}
The two transverse components can be conveniently combined into a complex number $\psi_{n} = m_{n,x} + i m_{n,y}$. These complex variables have the following equation of motion:
\begin{equation}
i (J + i \alpha J) \dot{\psi}_n 
= - A(\psi_{n-1} - 2 \psi_{n} + \psi_{n+1}).
\label{eq:spin-wave-Heisenberg-chain}
\end{equation}
Solutions of this equation have the form of waves $\delta m_n(t) = C e^{i k n a - i \omega t}$, where $a$ is the lattice period and $k$ is the wavenumber; the frequency is 
\begin{equation}
\omega 
	= \frac{2A(1-\cos{ka})}{J + i \alpha J}.
\end{equation}

Two things are worth noting, First, the frequency of a spin wave $\omega$ vanishes when the wavenumber $k \to 0$. This is a consequence of spontaneous breaking of a continuous symmetry, in this case of global spin rotations, by the ground state with spontaneous magnetization $\mathbf m$. Second, in the presence of dissipation ($\alpha > 0$), the frequency has an imaginary part,
\begin{equation}
\mathop{\mathrm{Im}{\omega}} 
	= - \frac{2 \alpha A(1-\cos{ka})}{J(1 + \alpha^2)}.
\end{equation}
Because $\mathop{\mathrm{Im}{\omega}} < 0$, a spin wave $\psi_n(t) = C e^{i k n a - i \omega t}$ decays exponentially in time. 

The mathematics can be simplified for waves whose wavelength is long compared to the lattice spacing, $k a \ll 1$. For such waves, magnetizations of adjacent dipoles are almost the same, $\mathbf m_{n+1} \approx \mathbf m_n$. In this limit, we may treat discrete variables $\mathbf m_n$ as a slowly varying and continuous function $\mathbf m(z)$, where $z = na$ is the coordinate along the chain. Then finite differences can be transformed into spatial derivatives via Taylor expansion:
\begin{eqnarray}
\mathbf m_{n+1} - \mathbf m_n 
	= \mathbf m(z+a) - \mathbf m(z) 
    \approx \frac{\partial \mathbf m(z)}{\partial z} a,
\nonumber\\
\mathbf m_{n+1} - 2\mathbf m_n + \mathbf m_{n-1}
    \approx \frac{\partial^2 \mathbf m(z)}{\partial z^2} a^2,
\end{eqnarray}
and so on.

In the continuum approximation, the equation of motion for transverse fluctuations $\psi(z,t) = m_x(z,t) + i m_y(z,t)$ (\ref{eq:spin-wave-Heisenberg-chain}) reads
\begin{equation}
i J \frac{\partial \psi}{\partial t} = - A a^2 \frac{\partial^2 \psi}{\partial z^2}. 
\end{equation}
For simplicity, we switched off dissipation, $\alpha \to 0$. The resulting wave equation resembles the Schr{\"o}dinger equation for a free particle with a nonrelativistic energy-momentum relation $E = p^2/2m$ and the mass $m = J \hbar/(2a^2A)$. These particles were first introduced by Felix Bloch \cite{Z.Phys.61.206} and are now known as magnons, the quanta of spin waves. 

\subsection{Continuum theory of a ferromagnet}

To develop a continuum description of a ferromagnetic medium, on a formal level, we promote the unit vector $\mathbf m$ encoding the magnetic moment of a single dipole to a unit-vector field $\mathbf m(\mathbf r)$ varying smoothly in space $\mathbf r$. The mathematical apparatus developed in the previous section is generalized accordingly. 

For example, the equation of motion (\ref{eq:2nd-law-spin-with-friction-recap}) undergoes two very minor changes: 
\begin{equation}
- \mathcal J \dot{\mathbf m} \times \mathbf m - \frac{\delta U}{\delta \mathbf m} - \alpha \mathcal J \dot{\mathbf m} = 0.
\label{eq:2nd-law-field-with-friction}
\end{equation}
First, the ``length'' of angular momentum $J$ has been replaced with the corresponding intensive quantity, the angular momentum $\mathcal J$ per unit volume. Second, the potential energy function $U(\mathbf m)$ has become the potential energy functional $U[\mathbf m(\mathbf r)]$ and thus the ordinary derivative $dU/d \mathbf m$ (\ref{eq:dU/dm-defined}) has turned into the functional derivative $\delta U/\delta \mathbf m(\mathbf r)$. 

The Landau-Lifshitz equations (\ref{eq:LLG-spin}) and (\ref{eq:LL-spin}) retain their forms, albeit with an effective field
\begin{equation}
\mathbf h_\mathrm{eff}(\mathbf r) = - \frac{1}{\mathcal M} \frac{\delta U}{\delta \mathbf m(\mathbf r)}
\end{equation}
containing the functional derivative. 

Equations of motion for the spherical-angle fields $\theta(\mathbf r)$ and $\phi(\mathbf r)$ are
\begin{eqnarray}
- \mathcal J \sin{\theta} \, \dot{\phi} 
	- \frac{\delta U}{\delta \theta} 
    - \alpha \mathcal J \dot{\theta} 
    = 0,
\nonumber\\
\mathcal J \dot{\theta} 
	- \frac{1}{\sin{\theta}} \frac{\delta U}{\delta \phi} 
    - \alpha \mathcal J \sin{\theta} \, \dot{\phi}
    = 0.
\label{eq:2nd-law-field-with-friction-spherical-angles}
\end{eqnarray}

A field $\mathbf m(\mathbf r)$ has infinitely many degrees of freedom, so if we attempt to describe it in terms of some coordinates $q^i$, the number of these coordinates will be infinite. Nonetheless, the equations of motion for these coordinates remain unchanged,
\begin{equation}
G_{ij} \dot{q}^j - \partial U/\partial q^i - D_{ij} \dot{q}^j = 0,
\label{eq:2nd-law-field-with-friction-general-variables}
\end{equation}
and only the gyroscopic and dissipative tensors are promoted to functionals of the magnetization field:
\begin{equation}
G_{ij} = - \mathcal J \int dV \, \mathbf m \cdot 
	\left( 
    	\frac{\partial \mathbf m}{\partial q^i} 
        	\times \frac{\partial \mathbf m}{\partial q^j} 
    \right), 
\quad
D_{ij} = \alpha \mathcal J \int dV \, 
	\frac{\partial \mathbf m}{\partial q^i}
    	\cdot \frac{\partial \mathbf m}{\partial q^j}.
\label{eq:G-and-D-field}
\end{equation}
The gyroscopic tensor can be obtained from a gauge potential $A_i$: 
\begin{equation}
G_{ij} 
= \frac{\partial A_j}{\partial q^i} 
- \frac{\partial A_i}{\partial q^j},
\quad
A_i = \int dV \, \mathbf a(\mmm) 
    \cdot \frac{\partial \mmm}{\partial q^i}.
\label{eq:G-via-A-soliton}
\end{equation}
This is entirely analogous to Eq.~(\ref{eq:G-via-A-spin}) for a single spin. Equations of motion (\ref{eq:2nd-law-field-with-friction-general-variables}) can be obtained from the Lagrangian 
\begin{equation}
L = A_i(q) \dot{q}^i - U(q)    
\label{eq:L-soliton}
\end{equation}
supplemented by the Rayleigh dissipation function \cite{LL-I}
\begin{equation}
R = \frac{1}{2} D_{ij} \dot{q}^i \dot{q}^j.    
\end{equation}

Our model is a ferromagnet in one spatial dimension with coordinate $z$. It can describe a ferromagnetic wire whose transverse dimensions are smaller than the characteristic length $\lambda_0$ defined in Eq.~(\ref{eq:units-easy-axis}).  Exchange interaction tends to align spins, so a ground state should have a uniform magnetization field $\mathbf m(z) = \mathrm{const}$. However, some of these uniform states will have lower energies than others. In particular, long-range dipolar interactions prefer that the dipoles line up along the direction of the wire, leaving just two ground states,
\begin{equation}
\mathbf m(z) = \pm \mathbf e_z = (0,0,\pm 1).
\label{eq:wire-vacua}
\end{equation}

\subsection{Model with local interactions}

We will use a simple phenomenenological form of the potential energy functional minimized by these two ground states: 
\begin{equation}
U[\mathbf m(z)] = \int dz 
	\left( 
    	A |\mathbf m'|^2 
        + K |\mathbf m \times \mathbf e_z|^2
    \right)/2.
\label{eq:U-wire-m}
\end{equation}
Here $\mathbf m' \equiv d\mathbf m/dz$. The first term models the exchange interaction and penalizes spatially non-uniform states. It can be derived as a continuum approximation to the Heisenberg chain. The second term penalizes states with magnetization deviating from the direction of the wire $\mathbf e_z$. 

Both the exchange constant $A$ and anisotropy strength $K$ are positive. Together with the density of angular momentum $\mathcal J$, they define the characteristic scales of length $\lambda_0$, time $\tau_0$, and energy $\epsilon_0$: 
\begin{equation}
\lambda_0 = \sqrt{A/K}, 
\quad
\tau_0 = \mathcal J/K,
\quad
\epsilon_0 = \sqrt{AK}.
\label{eq:units-easy-axis}
\end{equation}

We stress that this model is not very realistic as it relies on a local anisotropy term to select the two ground states (\ref{eq:wire-vacua}). A more realistic model with dipolar interactions would require a non-local energy functional as spins would interact over large distances. The advantage of our toy model is in its relative simplicity: with not too much effort, we can study its linear and nonlinear excitations. 

\subsection{Linear spin waves}

To obtain spin waves near one of the uniform ground states, $\mathbf m_0 = \mathbf e_z$, we consider a weakly excited state with $\mathbf m(z) = \mathbf m_0 + \delta \mathbf m(z)$, where $\delta \mathbf m = \mathbf e_x m_x + \mathbf e_y m_y$ represents small transverse deviations of magnetization. Expand the Lagrangian density to the second order in $\delta \mathbf m$ as in Sec.~\ref{sec:spin-precession-near-minimum}:  
\begin{equation}
L = 
	\frac{1}{2} \int dz \, 
    \left(
		\dot{m}_x m_y 
        	- \dot{m}_y m_x 
        - {m_x'}^2 - {m_y'}^2 
        - m_x^2 - m_y^2
	\right) + \ldots
\label{eq:L-linear-spin-waves}
\end{equation}
Here we have switched to natural units (\ref{eq:units-easy-axis}), in which $\mathcal J = A = K = 1$. The Lagrangian (\ref{eq:L-linear-spin-waves}) is quadratic in the transverse spin components, The resulting equations are linear in them, hence the term \emph{linear spin waves}. 

In the absence of dissipation, $\alpha=0$, the equations of motion for the transverse magnetization components are
\begin{equation}
- \dot{m}_y + m_x'' - m_x = 0,
\quad
\dot{m}_x + m_y'' - m_y = 0.
\end{equation}
It is convenient to combine the two transverse components into a complex field $\psi = m_x + i m_y$, whose equation of motion,
\begin{equation}
i \dot{\psi} = - \psi'' + \psi,
\end{equation}
resembles the Schr{\"o}dinger equation $i \dot{\psi} = \mathcal H \psi$ with the Hamiltonian 
\begin{equation}
\mathcal H = -\frac{d^2}{dz^2} + 1,
\label{eq:H-SUSY-flat}
\end{equation}
whose energy spectrum is $\epsilon_k = 1 + k^2$. A traveling spin wave $\psi(t,z) = \psi(0,0) e^{-i \omega t + i k z}$ has the frequency 
\begin{equation}
\omega = \frac{K + A k^2}{\mathcal J},
\label{eq:omega-k-linear}
\end{equation}
where we have restored the units. The frequency spectrum has a gap with a minimal excitation frequency $\lim_{k \to 0}\omega = K/\mathcal J = 1/\tau_0$. 

\subsection{Nonlinear spin waves}

\begin{figure}
    \centering
    \includegraphics[width=0.5\columnwidth]{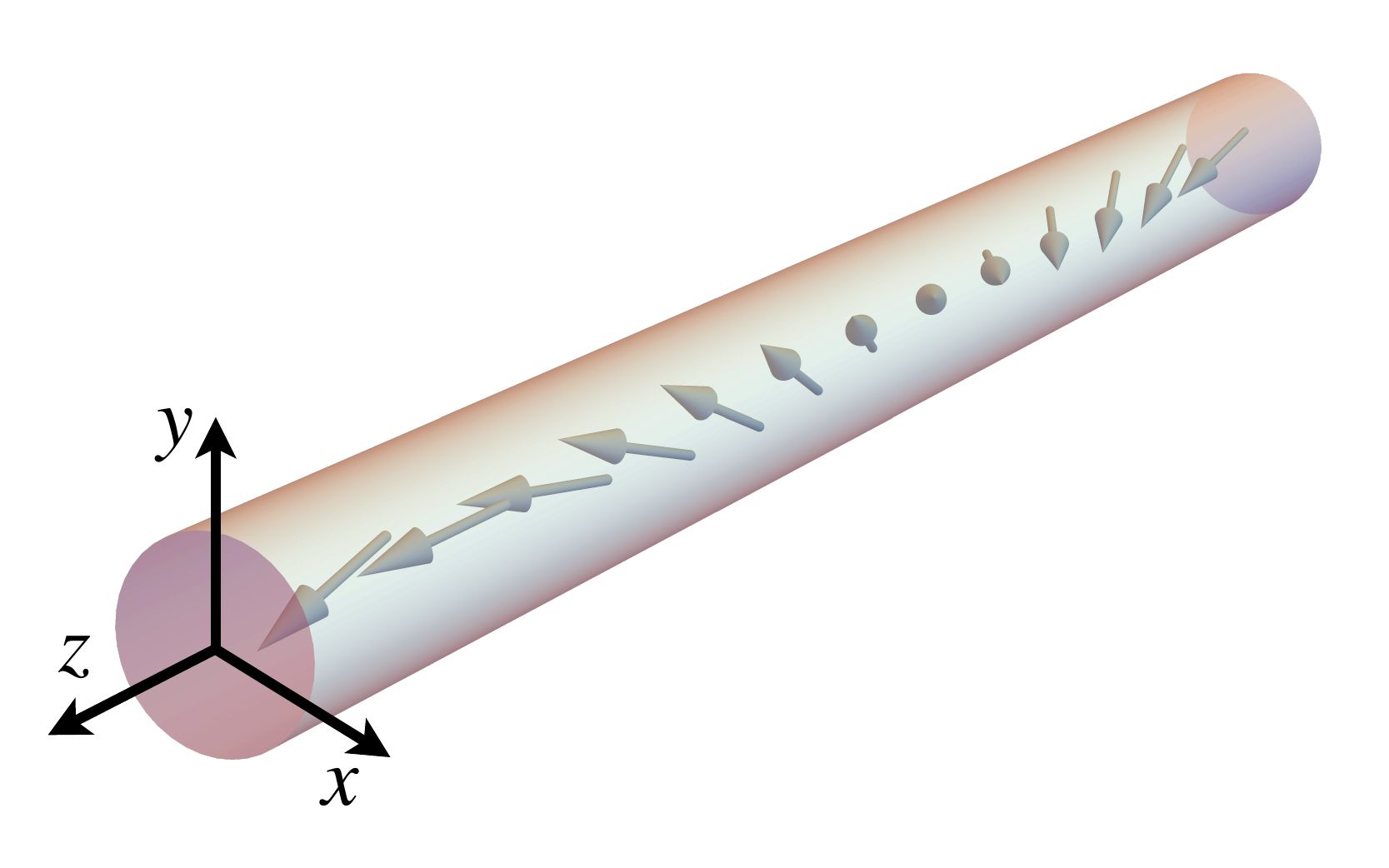}
    \caption{Snapshot of a large-amplitude spin wave. The spins have a fixed polar angle $\Theta$. The azimuthal angle varies in space as $\phi = kz + \Phi$.}
    \label{fig:spin-wave}
\end{figure}

To describe spin waves with a large amplitude, we introduce an Ansatz 
\begin{equation}
\mmm(z,t) = 
(\sin{\Theta} \cos{(kz + \Phi)}, \sin{\Theta} \sin{(kz + \Phi)}, \cos{\Theta}).
\label{eq:Ansatz-spin-wave}
\end{equation}
The polar angle $\Theta$ between $\mmm$ and the easy axis $\mathbf e_z$ is the wave's amplitude; the azimuthal $\Phi$ is its phase, Fig.~\ref{fig:spin-wave}. We will treat the amplitude $\Theta(t)$ and the phase $\Phi(t)$ as two collective coordinates of the nonlinear wave (\ref{eq:Ansatz-spin-wave}). Eq.~(\ref{eq:2nd-law-field-with-friction-general-variables}) determines their dynamics. 

The energy of the wave (\ref{eq:Ansatz-spin-wave}) scales linearly with the wire length and diverges in an infinite wire. It is therefore convenient to assume a finite wire length $\ell$ and to impose periodic boundary conditions to eliminate edge effects. The wire length $\ell$ must then be commensurate with the wavelength $\lambda = 2\pi/k$. The energy of the wave (\ref{eq:Ansatz-spin-wave}) is then 
\begin{equation}
U(\Theta,\Phi) = \frac{1}{2} (A k^2 + K) \ell \sin^2{\Theta}.    
\end{equation}
The lack of a $\Phi$ dependence is related to a symmetry, in this case the symmetry of global spin rotations about the $z$ axis. We say that $\Phi$ is a \emph{zero mode}. 

The gyroscopic and dissipative tensors are obtained with the aid of Eq.~(\ref{eq:G-and-D-field}). Their nonvanishing coefficients are 
\begin{equation}
G_{\Theta\Phi}= - G_{\Phi\Theta} 
= - \mathcal J \ell \sin{\Theta},  \quad
D_{\Theta\Theta} = \alpha \mathcal J \ell, 
\quad
D_{\Phi\Phi} = \alpha \mathcal J \ell \sin^2{\Theta}.
\end{equation}
The following equations of motion result for the collective coordinates: 
\begin{eqnarray}
\Theta: &&  
- \mathcal J \ell \sin{\Theta} \, \dot{\Phi}
- \frac{1}{2}(K + A k^2) \ell \sin{2\Theta}
- \alpha \mathcal J \ell \, \dot{\Theta} = 0,
\nonumber\\
\Phi: && + \mathcal J \ell \sin{\Theta} \, \dot{\Theta}
 - \alpha \mathcal J \ell \sin^2{\Theta} \, \dot{\Phi} = 0.
\end{eqnarray}
The equation for $\Phi$ shows that $\Theta$ is the slower variable of the two as $\dot{\Theta} = \alpha \sin{\Theta} \, \dot{\Phi}$ and $\alpha \ll 1$. It would be conserved in the absence of dissipation ($\alpha=0$). After some simplification, we find that the spin rotation frequency depends on the amplitude $\Theta$: 
\begin{equation}
\dot{\Phi} = - \frac{K + Ak^2}{\mathcal J(1+\alpha^2)} \cos{\Theta}.    
\label{eq:omega-k-nonlinear}
\end{equation}
Eq.~(\ref{eq:omega-k-nonlinear}) reproduces the spectrum of linear spin waves (\ref{eq:omega-k-linear}) in the limit of a small amplitude ($\Theta \to 0$) and weak dissipation ($\alpha \to 0$).

\subsection{Static domain wall}

The uniform ground states $\mathbf m(z) = \pm \mathbf e_z$ are global minima of the potential energy (\ref{eq:U-wire-m}). There are also local minima, in which magnetization interpolates between $-\mathbf e_z$ on one end of the wire and $+\mathbf e_z$ on the other. In such a state, magnetization $\mathbf m(z)$ forms two (nearly) uniform domains, in which the energy density is (nearly) zero. The region in between has non-uniform magnetization deviating from the easy axis has a positive energy density and is known as a domain wall. 

\begin{figure}
\centering
\includegraphics[width=0.9\columnwidth]{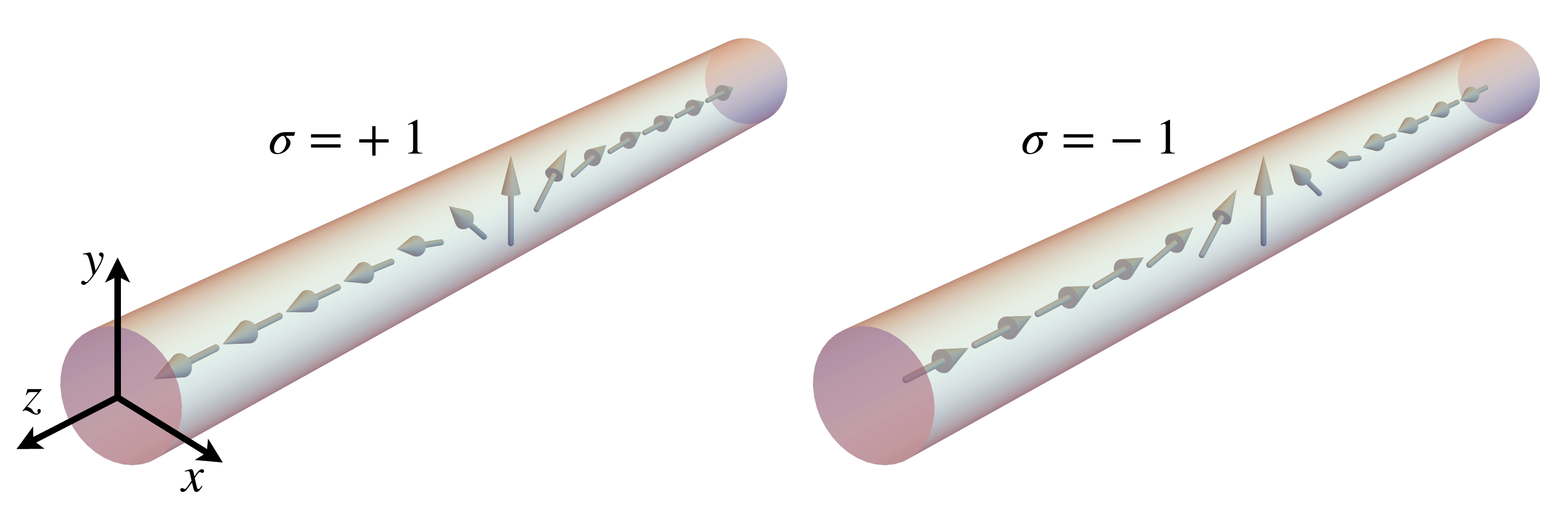}
\caption{Domain walls in a ferromagnetic wire with $Z_2$ topological charges $\sigma=\pm1$. }
\label{fig:domain-wall-1d}
\end{figure}

To obtain a formal solution for a domain wall, it is convenient to write the potential energy (\ref{eq:U-wire-m}) in terms of spherical angles, 
\begin{equation}
U[\theta(z),\phi(z)] = \int dz 
	\left[
    	A({\theta'}^2 + \sin^2{\theta} \, {\phi'}^2)
        + K \sin^2{\theta}
    \right]/2.
\label{eq:U-wire-spherical-angles}
\end{equation}
Minimization of the energy $\delta U = 0$ yields field equations 
\begin{equation}
-A \theta'' + \frac{1}{2}(A {\phi'}^2 + K)\sin{2\theta} = 0,
\quad
A(\sin^2{\theta} \, \phi')' = 0.
\end{equation}
Solutions with a uniform azimuthal angle, $\phi' = 0$ automatically satisfy the second of these equations. The first one can be integrated once to obtain $- A {\theta'}^2 + K \sin^2{\theta} = C$. The constant of integration can be obtained by noting that at $z \to \pm \infty$ we recover uniform states with $\theta' = 0$ and $\sin{\theta} = 0$. Therefore $C=0$. Integrating this equation yields domain-wall solutions (Fig.~\ref{fig:domain-wall-1d}), 
\begin{equation}
\cos{\theta(z)} = \sigma \tanh{\frac{z-Z}{\lambda_0}}, 
\quad 
\phi(z) = \Phi.
\label{eq:dw-static}
\end{equation}

Solutions (\ref{eq:dw-static}) have three parameters, one discrete and two continuous. 
\begin{itemize}
\item 
Topological charge $\sigma = \pm 1$ determines the ground states on the two sides of the domain wall: $\mathbf m = - \sigma \mathbf e_z$ at $z = -\infty$ and $\mathbf m = \sigma \mathbf e_z$ at $z = +\infty$. The word ``topological'' indicates that this characteristic cannot be changed by small continuous deformations of the soliton. 
\item
Position $Z$ determines the location of the domain wall. The change from $\mathbf m = - \sigma \mathbf e_z$ to $\sigma \mathbf e_z$ happens around $z = Z$, in a region of width 
$\lambda_0$ (\ref{eq:units-easy-axis}).
\item
Angle $\Phi$ defines an azimuthal plane in which the spins interpolate between $\mathbf m = - \sigma \mathbf e_z$ and $\sigma \mathbf e_z$. 
\end{itemize}

Note that the domain-wall solution (\ref{eq:dw-static}) contains two parameters, $Z$ and $\Phi$, defining the position of the domain wall and its azimuthal plane. These \emph{collective coordinates} of a domain wall quantify its rigid translations and rotations, transformations that leave the energy (\ref{eq:U-wire-m}) invariant and thus known as \emph{zero modes}. 

It is helpful to picture the configuration space 
\begin{equation}
-\infty < Z < +\infty, 
\quad
0 \leq \Phi \leq 2\pi
\end{equation}
as the surface of a cylinder of the unit radius. The state of a domain wall is then depicted by a point $(\cos{\Phi}, \sin{\Phi}, Z)$ on that surface. We will use this representation to illustrate the dynamics of a domain wall in Walker's model in Sec.~\ref{sec:Walker}. 

\subsection{Domain wall in a magnetic field}
\label{sec:domain-wall-wire-magnetic-field}

A magnetic field applied along the easy axis, $\mathbf h = h \mathbf e_z$, exerts a force on a domain wall, as can be seen from the following consideration. The field couples to magnetization through the Zeeman term with energy density $- \mathcal M \mathbf h \cdot \mathbf m$, where the magnetization density is $\mathcal M = \gamma \mathcal J$. By moving the domain wall, we can reduce its Zeeman energy by increasing the length of the domain where $\mathbf m$ points along $\mathbf h$. Displacing the domain wall (\ref{eq:dw-static}) by $\Delta Z$ changes the Zeeman energy by $2 \sigma \mathcal M h \Delta Z$, which indicates that the applied field exerts force 
\begin{equation}
F_Z = - 2 \sigma \mathcal M h 
\label{eq:F-Z-domain-wall}
\end{equation}
on the wall. 

It is natural to expect that the domain wall will move in the direction of the applied force. However, we shall see shortly, that the primary effect of the force will be precession of the domain wall, rather than translational motion. In fact, in the absence of dissipation the domain wall will not move at all, $\dot{Z} = 0$, because conservation of energy precludes that. The presence of dissipation will allow translational motion with a velocity proportional to the damping constant $\alpha$. 

To determine the dynamics of the domain wall in a magnetic field, we use the equations of motion for the fields $\theta(z)$ and $\phi(z)$ (\ref{eq:2nd-law-field-with-friction-spherical-angles}) with the energy functional 
\begin{equation}
U = \int dz 
	\left\{
    	\frac{A}{2}({\theta'}^2 + \sin^2{\theta} \, {\phi'}^2)
        + \frac{K}{2} \sin^2{\theta}
    - \mathcal M h \cos{\theta}
    \right\}.
\end{equation}
The equations of motion are 
\begin{eqnarray}
- \mathcal J \sin{\theta} \, \dot{\phi}
	+ A \theta'' 
    - \frac{1}{2}(A {\phi'}^2 + K)\sin{2\theta} 
    - \mathcal M h \sin{\theta}
    - \alpha \mathcal J \dot{\theta}
    = 0,
\nonumber\\
	\mathcal J \dot{\theta}
	- A(\sin^2{\theta} \, \phi')'
    - \alpha \mathcal J \sin{\theta} \, \dot{\phi}
	= 0.
\label{eq:eom-dw-in-field}
\end{eqnarray}
Although this looks like a big mess, it turns out that our previous solution can still be made to work, with one minor modification. We simply assume that the collective coordinates $Z$ and $\Phi$ in Eq.~(\ref{eq:dw-static}) become dependent on time. Eqs.~(\ref{eq:eom-dw-in-field}) are then satisfied if the linear and angular velocities obey the relations 
\begin{equation}
- \mathcal J \dot{\Phi} 
	- \mathcal M h 
    -\sigma \alpha \mathcal J \dot{Z}/\lambda_0 = 0,
\quad
\sigma \mathcal J \dot{Z} 
	- \alpha \mathcal J \lambda_0 \dot{\Phi} = 0.
\label{eq:eom-domain-wall-H-exact}
\end{equation} 
We thus obtain the velocities
\begin{equation}
\dot{\Phi} = - \tilde{\omega}_L, 
\quad
\dot{Z} = -\sigma \alpha \lambda_0 \tilde{\omega}_L.
\end{equation}
Here $\tilde{\omega}_L$ is the Larmor precession frequency renormalized by dissipation defined in Eq.~(\ref{eq:precession-spin-with-friction}).  

Two things immediately stand out. First, as we have already mentioned, the primary effect of the applied force (\ref{eq:F-Z-domain-wall}) is precession: the angular velocity $\dot{\Phi}$ is of order $\alpha^0$, whereas the linear velocity $\dot{Z}$ is of order $\alpha^1$. Second, the precession frequency of a domain wall is the same as that of a single spin (\ref{eq:precession-spin-with-friction}). The complexity brought by interactions of spins has no effect on the precession frequency. Simple results call for simple explanations, but our purely mathematical approach to the problem---find a solution to coupled partial differential equations---obscures the physical picture. Furthermore, closed-form solutions are available only in a very restricted set of problems. It worked for a magnetic field along the easy axis but it would fail for a field orthogonal to it. We need to find another approach.

\subsection{Weakly perturbed domain wall}

Fortunately, there is a general approach that will provide a path to solution for all kinds of perturbations and will be provide more physical intuition. Of course, there is a price that we will pay for the universality of the approach: it is only guaranteed to work for weak perturbations. 

The method is based on rigidity of a domain wall (and solitons in general). It takes a finite amount of energy to alter the width of a domain wall from its equilibrium value $\lambda_0$. On the other hand, no energy is required to translate a domain wall or to rotate its azimuthal plane. One can make an analogy with a tennis ball that is hard to squeeze but easy to move. Thus we may expect that a weak enough perturbation will produce substantial response only in the zero modes $Z$ and $\Phi$ of a domain wall but not in other modes (such as its width $\lambda_0$). 

On a formal level, although a domain wall can be deformed in an infinite number of ways and is therefore described by infinitely many coordinates $q^i$, we only need to take into account two of them---position $Z$ and azimuthal angle $\Phi$---when we consider the motion of a domain wall under a weak external perturbation. 

Thus the recipe is to use the equations of motion (\ref{eq:2nd-law-field-with-friction-general-variables}) with $q^i = Z$ and $\Phi$ only. To that end, we need the gyrotropic and dissipative tensors $G_{ij}$ and $D_{ij}$ (\ref{eq:G-and-D-field}). The gyroscopic tensor $G_{ij}$ is antisymmetric, so with 2 coordinates we just need to compute one of its components, say, $G_{Z\Phi}$. The dissipative tensor $D_{ij}$ is symmetric and, for this choice of collective coordinates, diagonal; so it only has 2 nonzero components. 

We compute in detail only the gyroscopic coefficient
\begin{equation}
G_{Z\Phi} = - \mathcal J \int_{-\infty}^{\infty} dz \, 
	\mathbf m \cdot 
    	\left(
        	\frac{\partial \mathbf m}{\partial Z}
            \times 
            \frac{\partial \mathbf m}{\partial \Phi}
        \right).
\end{equation}
As $\mathbf m$ is a function of $z-Z$, we may replace $\partial \mathbf m/\partial Z$ with $- \partial \mathbf m/\partial z$. On the other hand, $\phi = \Phi$ everywhere, so 
\begin{equation}
\frac{\partial \mathbf m}{\partial \Phi} 
= \frac{\partial \mathbf m}{\partial \phi}
= \mathbf e_\phi \sin{\theta},
\end{equation}
where we used the local spin frame defined in Eq.~(\ref{eq:local-frame}) and Fig.~\ref{fig:local-frame}. 
Next, because only $\theta$ varies in space and $\phi$ does not, application of the chain rule yields
\begin{equation}
\frac{\partial \mathbf m}{\partial z} 
= \frac{d \theta}{d z} 
\frac{\partial \mathbf m}{\partial \theta}
= \frac{d \theta}{d z}
\mathbf e_\theta.
\end{equation}
With the aid of the identity $\mathbf m \cdot (\mathbf e_\theta \times \mathbf e_\phi) = 1$, we obtain
\begin{equation}
G_{Z\Phi} 
    = - \mathcal J \int_{-\infty}^{\infty} dz \, 
    	\frac{d \cos{\theta}}{d z}
    = - \mathcal J \cos{\theta(z)}\big|_{-\infty}^{+\infty}
    = - 2 \sigma \mathcal J.
\end{equation}
It is remarkable that the gyroscopic coefficient $G_{Z\Phi}$ depends just on the density of angular momentum $\mathcal J$ and the topological charge $\sigma$ but is completely insensitive to the precise shape $\theta(z-Z)$ of the domain wall! 

The dissipative coefficients are computed along similar lines: 
\begin{equation}
D_{ZZ} = 2\alpha \mathcal J /\lambda_0, 
\quad 
D_{\Phi\Phi} = 2\alpha \mathcal J \lambda_0,
\quad
D_{\Phi Z} = D_{Z \Phi} = 0
\end{equation}
They give the following the equations of motion for our collective coordinates: 
\begin{eqnarray}
Z: && - 2 \sigma \mathcal J \dot{\Phi} 
	+ F_Z - 2 \alpha \mathcal J \dot{Z}/\lambda_0 
    = 0, 
\nonumber\\
\Phi: && + 2 \sigma \mathcal J \dot{Z} 
	+ F_\Phi - 2 \alpha \mathcal J \lambda_0 \dot{\Phi}
    = 0,
\label{eq:eom-domain-wall-H-Z-Phi}
\end{eqnarray}
where $F_Z = - \partial U/\partial Z$ is conservative force and $F_\Phi = - \partial U/\partial \Phi$ conservative torque. 

Returning to the previously solved problem of a domain wall in an applied field parallel to the easy axis, we have a force $F_Z = - 2 \sigma \mathcal M h$ and torque $F_\Phi = 0$. With these, the new approximate equations of motion (\ref{eq:eom-domain-wall-H-Z-Phi}) reproduce the exact ones (\ref{eq:eom-domain-wall-H-exact}).

The new approach also provides an insight into the vexing question: why is the precession frequency of a domain wall the same as that of a single spin in the same magnetic field? In the absence of friction, $\alpha = 0$, the precession frequency can be obtained from the first equation (\ref{eq:eom-domain-wall-H-Z-Phi}): 
\begin{equation}
\dot{\Phi} 
	= - \frac{F_Z}{G_{Z\Phi}}
    = -\frac{2\sigma \mathcal M h}{2\sigma \mathcal J}
    = - \gamma h. 
\label{eq:Larmor-precession-domain-wall}
\end{equation}
Both the Zeeman force on the domain wall $F_Z = - 2\sigma \mathcal M h$ and the gyroscopic coefficient $G_{Z\Phi} = - 2 \sigma \mathcal J$ are insensitive to the detailed structure of the domain wall. Their ration determines the precession frequency. 

Equations of motion (\ref{eq:eom-domain-wall-H-Z-Phi}) also paint a simple physical picture of the mechanics of a domain wall in a ferromagnetic wire. With two degrees of freedom, position $Z$ and angle $\Phi$, it resembles a bead on a string, which can move along the string and rotate. However, the dynamics of the two objects are very different. For a bead, a force creates linear acceleration and a torque creates angular acceleration. For a domain wall, a force generates angular velocity, whereas a torque produces linear velocity. \emph{When pushed, a domain wall rotates, and when twisted, it moves!} 

We can also understand now how the addition of dissipation enables a domain wall to move under an external force. When it rotates (its primary response to a force), dissipation creates a dissipative torque directed against the direction of rotation. This torque causes linear motion. The proportionality of the dissipative torque to the damping constant $\alpha$ explains why the velocity of the soliton is also of order $\alpha$. 

\subsection{Normal modes of a domain wall}

So far we have approximated the dynamics of a domain wall by its two most prominent modes---position $Z$ and global azimuthal angle $\Phi$. We have justified the neglect of all other modes (such as the width of the domain wall $\lambda$) by their hard nature. In this subsection, we shall make a quantitative analysis of the hard modes and deduce time scales on which their dynamics can be neglected. These results were first obtained by Yan \emph{et al.} \cite{Yan:2011}. 

In natural units of length, time, and energy (\ref{eq:units-easy-axis}), the Lagrangian density is 
\begin{equation}
\mathcal L(\theta, \phi) = (\cos{\theta} - 1)\dot{\phi} 
	- ({\theta'}^2 + \sin^2{\theta} \, {\phi'}^2 + \sin^2{\theta})/2.
\end{equation}
A static domain wall with $\sigma = +1$, $Z = 0$ and $\Phi = 0$ is 
\begin{equation}
\cos{\theta_0(z)} = \tanh{z}, 
\quad
\phi_0(z) = 0.
\end{equation}

Expand the Lagrangian density in small deviations from this state $\delta \theta(z)$ and $\delta \phi(z)$ to the second order: 
\begin{equation}
\mathcal L(\theta_0 + \delta \theta, \phi_0 + \delta \phi)
	= \mathcal L_0 + \mathcal L_1 + \mathcal L_2 + \ldots,
\end{equation}
where $\mathcal L_0 = \mathcal L(\theta_0, \phi_0)$. The first-order term $\mathcal L_1 = (\cos{\theta_0} - 1) \delta \dot{\phi}$ is a time derivative; it thus does not contribute to the equations of motion and can be neglected. The second-order term is 
\begin{equation}
\mathcal L_2 
	= - \sin{\theta_0} \, \delta \theta \, \delta \dot{\phi} 
    - ({\delta \theta'}^2 + \sin^2{\theta_0} \, {\delta \phi'}^2 + \cos{2\theta_0} \, \delta \theta^2)/2
\end{equation}
Rewriting it in terms of transverse increments $\delta m_\theta$ and $\delta m_\phi$ (\ref{eq:m-theta-m-phi-def}) reveals a hidden $O(2)$ symmetry:
\begin{equation}
\mathcal L_2 = - \delta m_\theta \, \delta \dot{m}_\phi
	- (\delta m_\theta \, \mathcal H \, \delta m_\theta
        + \delta m_\phi \, \mathcal H \, \delta m_\phi)/2,
\end{equation}
where we have introduced the operator
\begin{equation}
\mathcal H = - \frac{d^2}{dz^2} + 1 - 2 \mathrm{\,sech}^2{z}.
\label{eq:H-Poschl-Teller}
\end{equation}
To utilize the symmetry, we switch to complex fields 
\begin{equation}
\psi 
	= \frac{\delta m_\theta + \, i \, \delta m_\phi}{\sqrt{2}},
\quad
\psi^* 
	= \frac{\delta m_\theta - i \, \delta m_\phi}{\sqrt{2}}.
\label{eq:psi-def}
\end{equation}
Their Lagrangian density 
\begin{equation}
\mathcal L_2 = i \psi^* \dot{\psi} 
	- \psi^* \mathcal H \psi
\end{equation}
yields the equations of motion
\begin{equation}
i \dot{\psi} = \mathcal H \psi, 
\quad
- i \dot{\psi}^* = \mathcal H \psi^*.
\end{equation}

The spin-wave ``Hamiltonian'' $\mathcal H$ (\ref{eq:H-Poschl-Teller}) is similar to its counterpart for the ground state (\ref{eq:H-SUSY-flat}). The main difference is the presence of an attractive P{\"o}schl--Teller potential well $\mathcal U(z) = -2\mathrm{\,sech}^2{z}$ \cite{Poeschl1933} in Eq.~(\ref{eq:H-Poschl-Teller}) caused by the presence of the domain wall. This Hamiltonian has remarkable properties. (See \ref{app:PT} for a brief review and Ref.~\cite{Cooper1995} for much more.) Its energy spectrum has a single bound state with exactly zero energy, $\mathcal H \Psi_0 = 0$, and a localized wavefunction 
\begin{equation}
\Psi_0(z) = 2^{-1/2} \mathrm{\,sech}{z}.
\label{eq:Psi0}
\end{equation}
Its continuum states are reflectionless and are in a one-to-one correspondence with the plane waves $e^{ikz}$ of the free spin-wave Hamiltonian (\ref{eq:H-SUSY-flat}):
\begin{equation}
\mathcal H \psi_k = \epsilon_k \psi_k,
\quad 
\epsilon_k = 1 + k^2,
\quad
\psi_k(z) 
	= \frac{\tanh{z} - i k}{\sqrt{1+k^2}} 
		e^{i k z}.
\label{eq:psi-k}
\end{equation}
Eigenstates with different $k$ are orthogonal to the zero mode $\Psi_0(z)$ and to one another,
\begin{equation}
\int dz \, \psi_{k'}^*(z) \psi_k(z) = 2\pi \delta(k-k'),
\end{equation}
The zero mode $\Psi_0(z)$ corresponds to the infinitesimal translations and rotations of the domain wall, whereas the free states $\psi_k(z)$ describe spin waves in the presence of the domain wall. 

An arbitrary infinitesimal deformation of the domain wall can be expanded in terms of the eigenmodes,
\begin{equation}
\psi(z,t) = Q_0(t) \Psi_0(z) 
	+ \int \frac{dk}{2\pi} \, q_k(t) \psi_k(z),
\end{equation}
Complex amplitudes $Q_0(t)$ and $q_k(t)$ form a complete set of collective coordinates for a domain wall. The amplitude of the zero mode is related to position and azimuthal angle,
\begin{equation}
Q_0 = Z + i \Phi,
\end{equation}
whereas $q_k$ are complex amplitudes of the spin waves (\ref{eq:psi-k}).

The Lagrangian of these collective coordinates is 
\begin{equation}
L_2 = \int dz \, \mathcal L_2
	= i Q_0^* \dot{Q}_0 
    + \int \frac{dk}{2\pi} \, 
    	\left(
        	i q_k^* \dot{q}_k 
            - \epsilon_k |q_k|^2
        \right).
\end{equation}
The Rayleigh function is 
\begin{equation}
R = \alpha |\dot{Q}_0|^2 
	+ \alpha \int \frac{dk}{2\pi} \, |\dot{q}_k|^2.
\end{equation}
From these we obtain the equations of motion:
\begin{equation}
i \dot{Q}_0 - \alpha \dot{Q}_0 = 0,
\quad
i \dot{q}_k - \epsilon_k q_k - \alpha \dot{q}_k = 0,
\end{equation}
which yields the eigenfrequencies
\begin{equation}
\Omega_0 = 0,
\quad
\omega_k = \frac{\epsilon_k}{1 + i \alpha} 
	= \frac{1+k^2}{1 + i \alpha}.
\end{equation}
Upon restoring the units, we obtain the spectrum
\begin{equation}
\omega_k = \frac{K + A k^2}{\mathcal J(1 + i \alpha {\mathcal J})}.
\label{eq:eigenfrequencies-domain-wall}
\end{equation}
The relaxation rate $\gamma_k = - \mathrm{Im}{\omega_k} \approx \alpha \epsilon_k$ for $\alpha \ll 1$ is slowest for modes with $k \to 0$, 
\begin{equation}
\lim_{k \to 0}{\gamma_k} = \alpha K/\mathcal J = \alpha/\tau_0.
\end{equation}
Thus the influence of spin waves can be neglected on time scales longer than $\tau_0/\alpha$.

This criterion can be further relaxed. If external perturbations and the response of the domain wall occur on time scales longer than $\tau_0$ then spin waves, whose spectrum has a gap $\omega_k \geq 1/\tau_0$, are not excited. 

\subsection{Angular momentum of a domain wall}

Symmetry with respect to rotations about the $z$ axis implies conservation of angular momentum $J_z$. In our one-dimensional model, there is no orbital contribution to this quantity. Therefore angular momentum comes solely from spin. This gives 
\begin{equation}
J_z(Z) = \int dz \, \mathcal J \mathbf m(z-Z) \cdot \mathbf e_z.
\end{equation}
This quantity is not well defined in an infinite wire, where the integration region extends from $z = -\infty$ to $+\infty$ because the integrand approaches the constant values $\pm \mathcal J$ away from the domain wall, $|z - Z| \gg \lambda$. It is convenient to consider a finite wire of finite length $\ell \gg \lambda$ so that the integration range is $-\ell/2 < z < \ell/2$. If the domain wall is located exactly in the middle of this wire, $Z = 0$, then the contributions of spins with positive and negative projections onto the $z$ axis cancel each other and we find $J_z = 0$. Shifting the domain wall to a position $Z \neq 0$ shifts the balance by elongating one domain and shortening the other, creating a net angular momentum proportional to the displacement of the domain wall, $J_z(Z) \propto Z$. To compute the proportionality constant, we may take the derivative 
\begin{eqnarray}
\frac{\partial J_z}{\partial Z} 
	&=& \mathcal J \int_{-\ell/2}^{+\ell/2} dz \,
		\frac{\partial \mathbf m(z-Z)}{\partial Z} \cdot \mathbf e_z
    = - \mathcal J \int_{-\ell/2}^{+\ell/2} dz \,
		\frac{\partial \mathbf m(z-Z)}{\partial z} \cdot \mathbf e_z
\nonumber\\
    &=& - \left. \mathcal J \mathbf m(z-Z) \cdot \mathbf e_z \right|_{-\ell/2}^{+\ell/2}
    \approx - 2 \sigma \mathcal J.
\end{eqnarray}
In the last step, we assumed that the magnetization at the ends of the wire returns to the ground-state values $\pm \sigma e_z$, which is justified if the domain wall is not too close to the ends. 

We thus obtain an important result: 
\begin{equation}
J_z(Z) = - 2 \sigma \mathcal J Z. 
\label{eq:angular-momentum-domain-wall}
\end{equation}
The angular momentum of a domain wall in a ferromagnetic wire is proportional to its position $Z$ relative to the center of the wire. The proportionality coefficient is insensitive to the detailed structure of the domain wall and depends only on the density of angular momentum $\mathcal J$ (per unit length) and the topological charge of the domain wall $\sigma$ defined in Eq.~(\ref{eq:dw-static}). 

This result has interesting consequences. Angle $\Phi$ and angular momentum $J_z$ form a canonical pair with the commutator $[\Phi,J_z] = i \hbar$. With the aid of Eq.~(\ref{eq:angular-momentum-domain-wall}), we find that coordinate $Z$ and azimuthal angle $\Phi$ do not commute, 
\begin{equation}
[\Phi,Z] = - \frac{i \hbar}{2 \sigma \mathcal J}.
\label{eq:commutator-Z-Phi}
\end{equation}
It follows that the position and azimuthal orientation of a domain wall cannot be simultaneously measured. The uncertainty product is $\Delta Z \Delta \Phi \geq \hbar / 4 \mathcal J $. In an atomic chain with the lattice constant $a$ and atomic spins of length $\hbar S$ (with $2S$ an integer), $\mathcal J = \hbar S/a$ and the uncertainty product $\Delta Z \Delta \Phi \geq a/4S$ is of the order of the interatomic distance. 

\subsection{Linear momentum of a domain wall}

In a uniform wire, translational symmetry implies conservation of linear momentum $P_z$. We cannot derive it along the lines leading to Eq.~(\ref{eq:angular-momentum-domain-wall}). Instead, we will reverse-engineer it from the commutator of $Z$ and $\Phi$ (\ref{eq:commutator-Z-Phi}). 

To that end, we recall that coordinate $Z$ and its momentum $P_z$ have the commutator $[P_z, Z] = - i \hbar$. Comparison with Eq.~(\ref{eq:commutator-Z-Phi}) shows that a plausible answer would be 
\begin{equation}
P_z = 2 \sigma \mathcal J \Phi.
\label{eq:linear-momentum-domain-wall}
\end{equation}
The linear momentum of a domain wall in a ferromagnetic wire is thus proportional to its azimuthal angle. 

Although this result is hard to internalize, we can check that it is compatible with the dynamics of a domain wall derived earlier. As we have seen previously, an external magnetic field $\mathbf h$ parallel to the easy axis $\mathbf e_z$ exerts on the domain wall the force $F_z = - 2 \sigma \mathcal M h$. This external force determines the rate of change for the linear momentum, $\dot{P}_z = - 2 \sigma \mathcal M h$. Our conjectured result for $P_z$ (\ref{eq:linear-momentum-domain-wall}) tells us that the domain wall will precess at the rate $\dot{\Phi} = - \gamma h$, in accordance with Eq.~(\ref{eq:Larmor-precession-domain-wall}).

Eqs.~(\ref{eq:angular-momentum-domain-wall}) and (\ref{eq:linear-momentum-domain-wall}) reveal a pleasing symmetry. The \emph{angular} momentum is proportional to the \emph{linear} coordinate of the domain wall and vice versa, with the same proportionality coefficient, up to a sign. 

There is, however, a subtle difference between the linear and angular coordinates: whereas the former lives on a straight line $-\infty < Z < +\infty$ and is unambiguously defined, the latter lives on a circle, so that values of $\Phi$ differing by $2\pi$ describe the same physical state and must be identified. This leads to an ambiguity in linear momentum (\ref{eq:linear-momentum-domain-wall}): it is defined modulo $4\pi \mathcal J$. There is no such problem with the angular momentum. 

The paradox of linear momentum was resolved by Haldane \cite{Haldane1986}. It turns out that the continuum and classical theory of the ferromagnet used throughout this section ``remembers'' its discrete and quantum origins. In a periodic atomic chain with the lattice constant $a$, linear momentum is defined modulo $2\pi \hbar/a$ thanks to Bragg scattering. If the atomic spins have length $\hbar S$, the spin density is $\mathcal J = \hbar S/a$ and the momentum ambiguity is $4\pi \hbar S/a$. This is indeed an integer multiple of $2\pi \hbar/a$ because $2S$ is an integer. 

\subsection{Counting states of a domain wall}

At sufficiently low energies (defined below), position $Z$ and angle $\Phi$ of a domain wall fully describe its dynamics. Since the two variables are, up to a multiplicative constant, momenta of one another, the low-energy configuration space $(Z,\Phi)$ can be viewed as a phase space, in which the number of states is proportional to the volume: 
\begin{equation}
d\Gamma = \frac{dZ \, dP_z}{2\pi \hbar}
	= \frac{d\Phi \, dJ_z}{2\pi \hbar}
    = \frac{\mathcal J \, dZ \, d\Phi}{\pi \hbar}.
\end{equation}

The same result can be obtained if we extrapolate Eq.~(\ref{eq:dGamma-spin}) from a single spin to a domain wall by using $q^1 = Z$, $q^2 = \Phi$, and $G_{12} = G_{Z\Phi} = - 2 \sigma \mathcal J$. 

In our count, we included the two zero modes of a domain wall and discarded all of the other modes, which (at low amplitudes) are spin waves with finite frequencies $\omega \geq \omega_0 = K/\mathcal J $ (\ref{eq:eigenfrequencies-domain-wall}). The spin waves are frozen out at energies below $\hbar \omega_0$. 

\subsection{Hamiltonian dynamics}

In Sec.~\ref{sec:spin-Hamilton}, we expressed the dynamics of a single spin in terms of canonical variables. There we chose the azimuthal angle $\phi$ as coordinate $q$ and the $z$ component of spin $J_z = J \cos{\theta}$ as canonical momentum $p$. This can also be done for a magnetic soliton. For a domain wall in a ferromagnetic wire, we also select the azimuthal angle $\Phi$ and angular momentum $J_z$ as a canonical pair: 
\begin{equation}
q = \Phi, 
\quad
p = - 2 \sigma \mathcal J Z,
\label{eq:p-q-domain-wall-wire}
\end{equation}
see Eq.~(\ref{eq:angular-momentum-domain-wall}). 

Let us consider a domain wall perturbed by a weak external magnetic field $\mathbf h$ parallel to the easy axis and small anisotropy breaking the axial symmetry. Then the energy of the domain wall will have the following form:
\begin{equation}
U(Z,\Phi) = 2\sigma \mathcal M h Z + V(\Phi). 
\end{equation}
The first term expresses the Zeeman coupling to the magnetic field (Sec.~\ref{sec:domain-wall-wire-magnetic-field}), the second is a phenomenological energy of the additional anisotropy. By expressing the energy in terms of the canonical variables (\ref{eq:p-q-domain-wall-wire}), we obtain the Hamiltonian
\begin{equation}
H(p,q) = - 2 \omega_L p + V(q).     
\end{equation}
Here $\omega_L$ is the Larmor precession frequency (\ref{eq:Larmor-frequency}).
Hamilton's equations (\ref{eq:Hamilton}) read 
\begin{equation}
\dot{q} = - \omega_L, 
\quad
\dot{p} = - d V/dq.
\end{equation}
It is straightforward to check that these equations reproduce the equations of motion (\ref{eq:eom-domain-wall-H-Z-Phi}) in the absence of dissipative forces, $\alpha=0$. 

The formulation of dynamics in terms of canonical variables underpins the standard procedure of quantization. With physical variables promoted to operators, the Poisson bracket turns into the commutator. Quantization of vortex and skyrmion dynamics along these lines was considered in Refs.~\cite{Galkin2007, Takashima2016}. 

It is worth noting that our experience with classical mechanics of massive particles and objects made from them has conditioned us to expect that canonical momenta are proportional to velocities: the linear momentum of a nonrelativistic particle is $p_x = m\dot{x}$ and the angular momentum of a planar rotator is $L = I \dot{\Phi}$. In contrast, the linear momentum of a domain wall in a wire is proportional to its azimuthal angle and the angular momentum to its coordinate, see Eqs.~(\ref{eq:angular-momentum-domain-wall}) and (\ref{eq:linear-momentum-domain-wall}). Roughly speaking, the two coordinates $Z$ and $\Phi$ of a domain wall also serve as its canonical momenta! Instead of two pairs of canonical variables in this case we have only one, so one coordinate has to serve as a canonical coordinate and the other as canonical momentum. 

For a ferromagnetic soliton, the separation of collective coordinates into canonical coordinates and momenta is arbitrary and therefore somewhat unsatisfactory. It can be avoided if we do not insist on a canonical structure and treat all collective coordinates on the same footing. In this treatment, the Poisson bracket can be obtained from its inverse known as the Lagrange bracket \cite{Goldstein2001}. For ferromagnetic solitons, the Lagrange bracket is the gyroscopic tensor. See Ref.~\cite{Gonzalez2022} for further details. 

\section{Walker's model of a domain wall}
\label{sec:Walker}

\begin{figure}
    \centering
    \includegraphics[width=0.6\columnwidth]{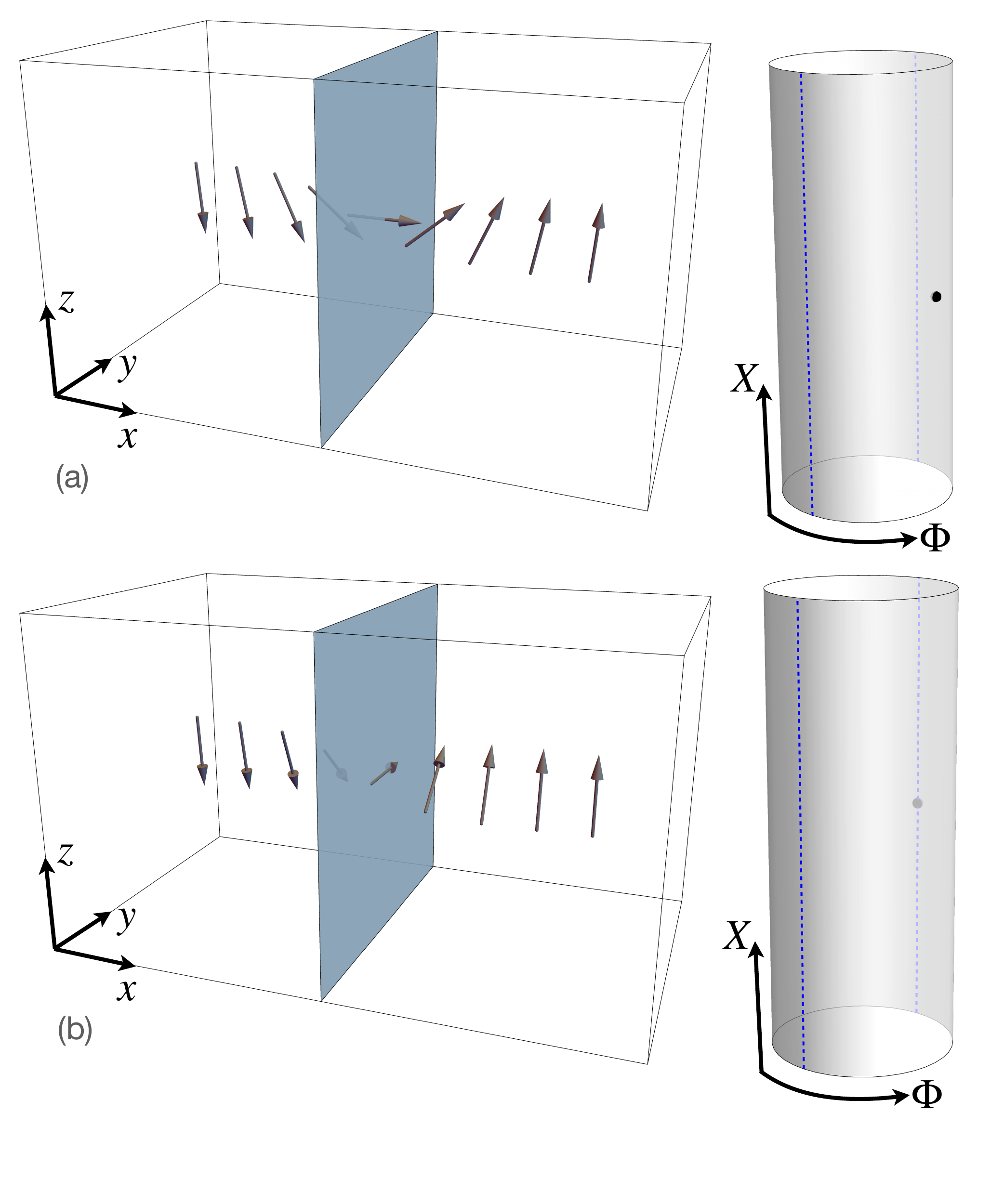}
    \caption{Domain wall in Walker's model: (a) $\Phi = 0$. (b) $\Phi = \pi/2$. The shaded plane is the plane of the domain wall, where $\mmm(x)$ is orthogonal to the easy axis $\mathbf e_z$. The dotted blue lines are minima of the anisotropy energy at $\Phi = \pm \pi/2$. The black dot on the surface of the cylinder depicts the configuration $(X,\Phi)$ of the domain wall.}
    \label{fig:domain-wall-walker}
\end{figure}

Schryer and Walker \cite{Schryer1974} investigated the dynamics of a realistic domain wall in a 3-dimensional ferromagnet with intrinsic easy-axis anisotropy. In their model, the easy direction is $z$ and the domain wall forms a sheet parallel to the $yz$ plane, Fig.~\ref{fig:domain-wall-walker}, so that the spin field depends on the $x$ coordinate only. The axial symmetry of global rotations about the $z$ axis encountered in the previous model of a ferromagnetic wire is broken by magnetostatic (dipolar) interactions and the energy of the domain wall depends on its azimuthal orientation. 

\subsection{Energy functional and domain-wall configuration}

The magnetization field $\mmm(x)$ has the following energy per unit area in Walker's model: 
\begin{equation}
U_0 = \int dx 
	\left[
    	\frac{A}{2}
        	({\theta'}^2 + \sin^2{\theta} {\phi'}^2)
        + \frac{K}{2} \sin^2{\theta}
        + \frac{\mathcal M^2}{8\pi}  \sin^2{\theta} \cos^2{\phi}
    \right].
\label{eq:U-Walker-spherical-angles}
\end{equation}
Here $\mathcal M$ is the magnetization related to the spin density $\mathcal J$ by the gyroscopic ratio, $\mathcal M = \gamma \mathcal J$. The first two terms in Eq.~(\ref{eq:U-Walker-spherical-angles}) are similar to those of our simplified 1-dimensional model (\ref{eq:U-wire-spherical-angles}); $A$ is the strength of Heisenberg exchange and $K$ is the easy-axis anisotropy. The last, magnetostatic term represents the energy of the magnetic field $\mathbf h = (-\mathcal M \sin{\theta} \cos{\phi}, 0, 0)$ generated by the inhomogeneous magnetization on the domain wall and can be viewed as additional anisotropy breaking the symmetry of rotations about the easy $z$-axis. The $x$ direction is made harder by the magnetostatic energy. Therefore, in equilibrium the magnetization will stay in the $yz$ plane. 

The characteristic length, time, and energy scales are the same as in the model of the axially symmetric wire (Sec.~\ref{sec:ferromagnetic-wire}):
\begin{equation}
\lambda_0 = \sqrt{A/K}, 
\quad
\tau_0 = \mathcal J/K,
\quad
\epsilon_0 = \sqrt{AK},
\quad
\mu^2 = \mathcal M^2/4\pi K.
\label{eq:units-Walker}
\end{equation}
The dimensionless ratio $\mu^2$ quantifies the relative strengths of the magnetostatic and intrinsic anisotropy terms. 

Minimizing the energy functional (\ref{eq:U-Walker-spherical-angles}) at a fixed and uniform azimuthal angle, $\phi(x) = \Phi$, yields a  profile shown in Fig.~\ref{fig:domain-wall-walker},
\begin{equation}
\cos{\theta(x)} = \sigma \tanh{\frac{x-X}{\lambda(\Phi)}},
\quad
\lambda(\Phi) 
= \frac{\lambda_0}{\sqrt{1+\mu^2 \cos^2{\Phi}}}.
\end{equation}
The energy density (per unit area) of a domain wall,
\begin{equation}
U_0(\Phi) = \epsilon_0\sqrt{1+\mu^2 \cos^2{\Phi}},
\end{equation}
is the lowest for $\Phi = \pm \pi/2$, i.e., when the domain wall interpolates the ground states $\mmm = -\mathbf e_z$ and $\mmm = +\mathbf e_z$ in the $yz$ plane. 

As in the model of a ferromagnetic wire, we will add an external perturbation in the form of the Zeeman coupling to a magnetic field $\mathbf h = (0,0,h)$ aligned with the easy axis,
\begin{equation}
U_1(X) = - \int dx \, \mathcal M h \cos{\theta} 
= 2 \sigma \mathcal M h X.
\end{equation}

Position $X$ and angle $\Phi$ are the collective coordinates of Walker's domain wall. It will be useful to picture the state of the domain wall in configuration space 
\begin{equation}
-\infty < X < +\infty,
\quad
0 \leq \Phi \leq 2\pi,
\end{equation}
the surface of a cylinder, Fig.~\ref{fig:domain-wall-walker}. The width of the domain wall $\lambda(\Phi)$ is a slave of the angle $\Phi$, not an independent variable. 

The gyrotropic and dissipative tensors have the following nonvanishing coefficients:
\begin{eqnarray}
G_{X\Phi} = - G_{\Phi X} = - 2 \sigma \mathcal J,
\nonumber\\
D_{XX} = \frac{2 \alpha \mathcal J}{\lambda(\Phi)},
\quad
D_{\Phi\Phi} = 2 \alpha \mathcal J \lambda(\Phi)
	\left[
    	1 + \left(
                \frac{\lambda'(\Phi)}{\lambda(\Phi)}
        	\right)^2
    \right].
\end{eqnarray}
Observe that the dissipative coefficients depend on the width of the domain wall $\lambda$ (which in turn depends on the azimuthal angle $\Phi$), indicating that the viscous force depends on the precise \emph{geometry} of the domain wall (in this case, its width). In contrast, the gyroscopic coefficients are not sensitive to the geometry and depend only on the \emph{topology} of the domain wall (through the $Z_2$ topological charge $\sigma$). 

\subsection{Dynamics of collective coordinates}

The motion of a domain wall in Walker's model is more complex than in the simpler model of a magnetic wire considered above. The increased complexity can be traced to a more elaborate energy landscape of the domain wall: its energy now depends on both collective coordinates,
\begin{equation}
U(X,\Phi) = 2 \sigma \mathcal M h X + U_0(\Phi),  
\label{eq:Walker-energy}
\end{equation}
so there are both conservative force and torque (per unit area): 
\begin{eqnarray}
F_X &=& - \frac{\partial U}{\partial X} 
= - 2 \sigma \mathcal M h, 
\nonumber\\
F_\Phi &=& - \frac{\partial U}{\partial \Phi}
= \frac{\epsilon_0\mu^2\sin{2\Phi}}{2\sqrt{1+\mu^2 \cos^2{\Phi}}}.
\label{eq:Walker-conservative-force-torque}
\end{eqnarray}
The force grows linearly with the applied field $h$. The torque attains a peak value $F_\Phi^\mathrm{max}$ at a critical angle $\Phi_c$, where
\begin{equation}
F_\Phi^\mathrm{max} 
=  \epsilon_0
\left(\sqrt{1+\mu^2}-1\right),
\quad
\Phi_c = \arctan{\sqrt[4]{1+\mu^2}}.
\label{eq:Walker-peak-torque}
\end{equation}

The equations of motion for the collective coordinates read
\begin{eqnarray}
X: && 
- 2 \sigma \mathcal J \dot{\Phi} + F_X - D_{XX} \dot{X} = 0,
\nonumber\\
\Phi: &&
+ 2 \sigma \mathcal J \dot{X} + F_\Phi - D_{\Phi\Phi} \dot{\Phi} = 0.
\label{eq:eom-Walker}
\end{eqnarray}
In the absence of the external field, $H=0$, the domain wall is at rest, $\dot{X} = 0$, at one of its equilibrium azimuthal orientations, $\Phi = \pm\pi/2$. Turning on the field sets the domain wall in motion. Its dynamics has two distinct regimes depending on the strength of the applied field $h$. 

\subsection{Steady-state motion in a weak field} 
\label{sec:Walker-steady-state}

Below a critical field strength $h_c$, the domain wall moves at a constant velocity, $\dot{X} = \mathrm{const}$, without changing its shape, $\dot{\Phi} = 0$. Its velocity grows with an increasing field, 
\begin{equation}
\dot{X} = \frac{F_X}{D_{XX}} 
= - \frac{\sigma \gamma h \lambda(\Phi)}{\alpha}. 
\label{eq:Walked-steady-state-velocity}
\end{equation}

As the velocity grows, the azimuthal angle $\Phi$ deviates more from the equilibrium value $\Phi = \pm\pi/2$. The azimuthal angle in the state of dynamical equilibrium is found from the balance of torques, given by the second of Eqs.~(\ref{eq:eom-Walker}),
\begin{equation}
2\sigma \dot{X} + \frac{\gamma \mathcal M \lambda(\Phi)}{4\pi}\sin{2\Phi} = 0.
\label{eq:Walker-balance-torques}
\end{equation}

\begin{figure}
    \centering
    \includegraphics[width=0.9\columnwidth]{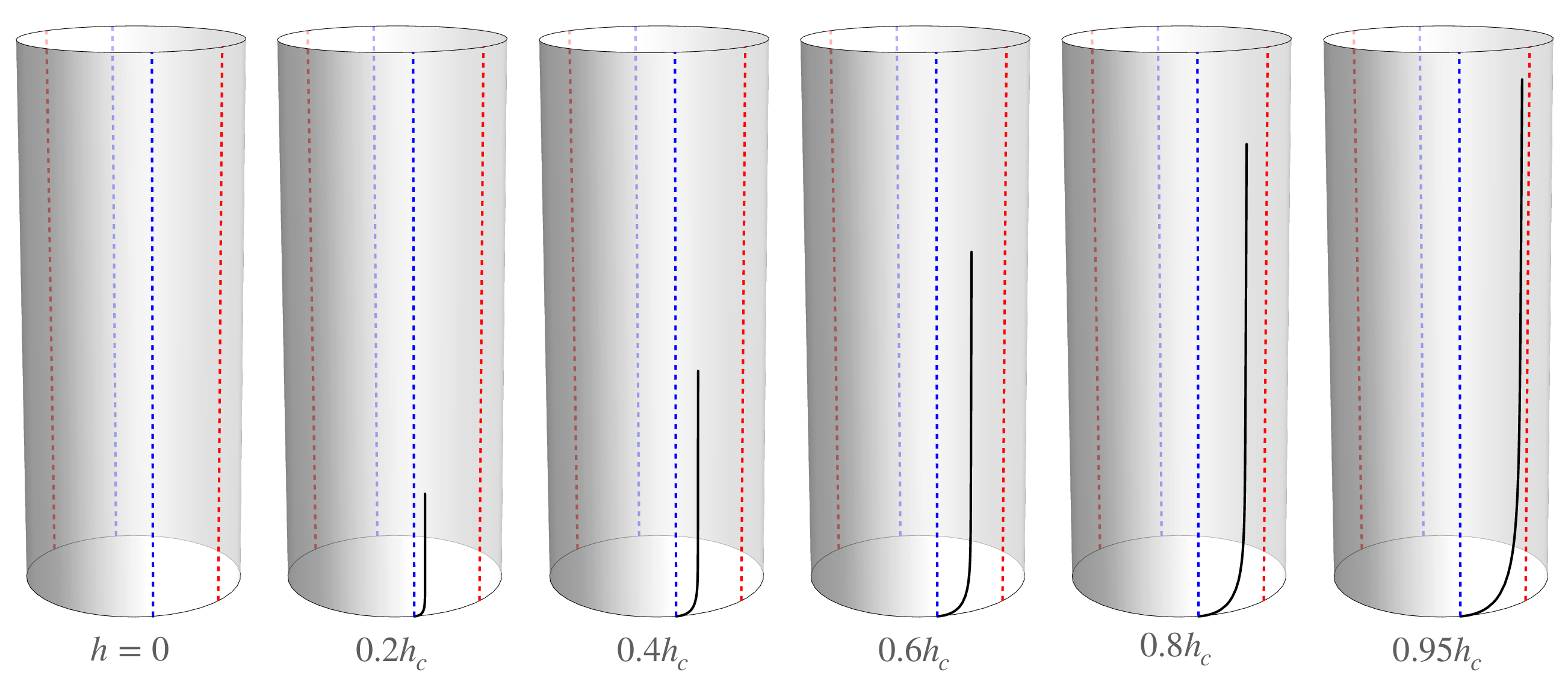}
    \caption{Motion of Walker's domain wall in weak fields, $h < h_c$. Numerical solutions of Eqs.~(\ref{eq:eom-Walker}) for the anisotropy ratio $\mu^2=1$ and Gilbert damping $\alpha=0.05$ Solid black line: trajectory of the domain wall in configuration space $(X,\Phi)$ depicted as the surface of a cylinder, as in Fig.~\ref{fig:domain-wall-walker}. Dashed blue lines: minima of the anisotropy energy $U_0(\Phi)$. Dashed red lines: critical angles $\Phi_c$. Trajectories for all fields $h$ are shown for the same length of time $2T_c$, where $T_c = 2\pi/(\gamma h_c)$ is the Larmor period for the critical field. The vertical displacement is an indicator of the domain wall velocity. }
    \label{fig:Walker-trajectory1}
\end{figure}

Fig.~\ref{fig:Walker-trajectory1} shows the motion of a domain wall through its configuration space $(X,\Phi)$ for weak fields $h<h_c$. The domain wall starts from a minimum of the anisotropy energy $U_0(\Phi)$ at $\Phi = \pi/2$ and reaches a new dynamic equilibrium at an angle $\Phi < \Phi_c$. All diagrams show the motion for the same length of time $2T_c$, where $T_c = 2\pi/(\gamma h_c)$ is the Larmor period for the critical field $h_c$. Thus the axial (vertical) displacement serves as a proxy for the linear velocity $\dot{X}$. As can be seen from Fig.~\ref{fig:Walker-trajectory1}, the steady-state velocity increases with an increasing field.

A steady state can be maintained as long as the gyroscopic torque $2\sigma \mathcal J \dot{X}$ is balanced by the conservative torque $F_\Phi$. The peak torque (\ref{eq:Walker-peak-torque}) determines the critical speed 
\begin{equation}
|\dot{X}|_c = \frac{F_\Phi^\mathrm{max}}{2\mathcal J}, 
\label{eq:Walker-critical-velocity}
\end{equation}
above which the steady-state motion cannot be sustained (Walker's breakdown). The critical field is 
\begin{equation}
h_c 
= \frac{\alpha \mathcal M}{16\pi} \sin{2\Phi_c}.
\label{eq:Walker-critical-field}
\end{equation}

Note that the critical velocity (\ref{eq:Walker-critical-velocity}) is independent of Gilbert's damping $\alpha$, whereas the critical field (\ref{eq:Walker-critical-field}) is of order $\alpha$. 

\subsection{Walker's breakdown}

\begin{figure}
    \centering
    \includegraphics[width=0.9\columnwidth]{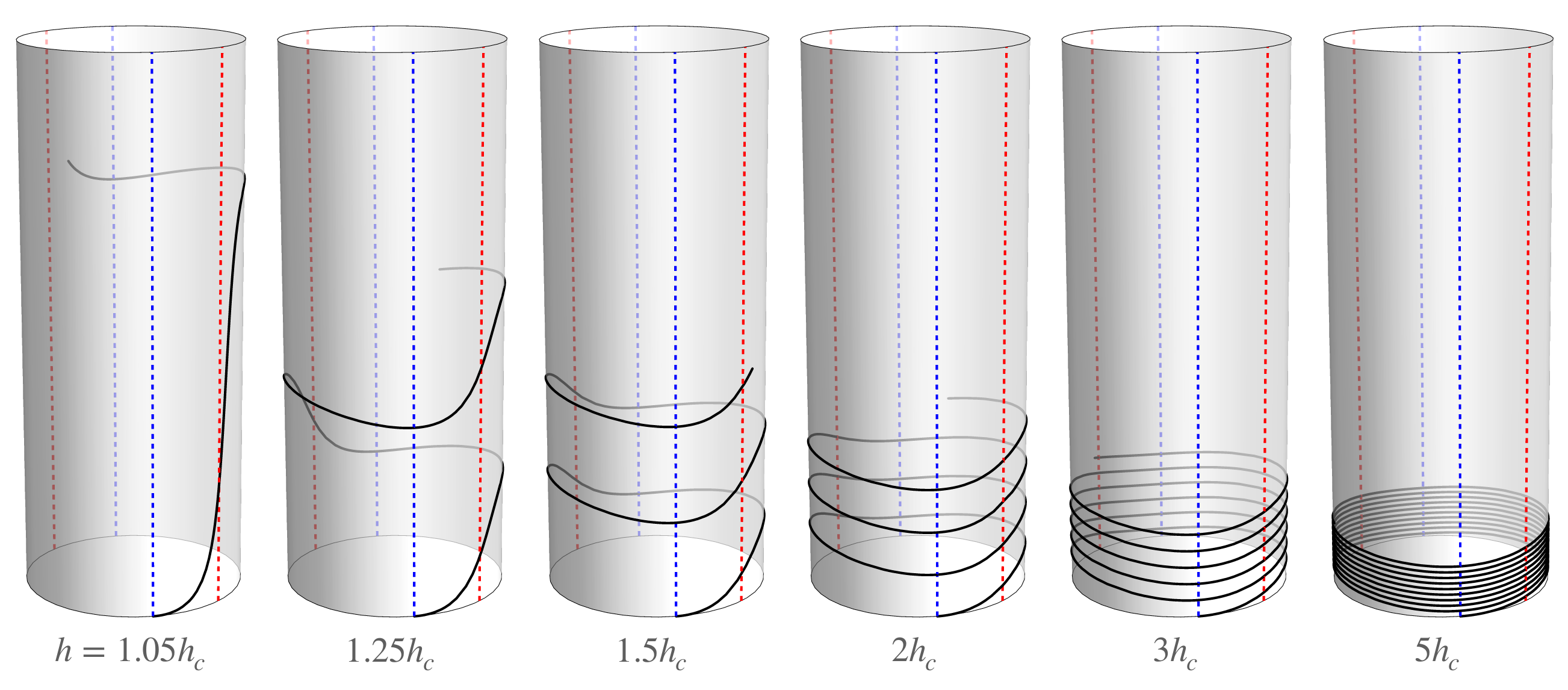}
    \caption{Motion of Walker's domain wall in strong fields, $h > h_c$. See Fig.~\ref{fig:Walker-trajectory1} for details. The shrinking axial displacement over the same time period $2T_c$ indicates a decreasing average velocity of the domain wall in higher fields.}
    \label{fig:Walker-trajectory2}
\end{figure}

Above the critical field, the steady-state velocity (\ref{eq:Walked-steady-state-velocity}) becomes so large that the gyroscopic torque $G_{\Phi X} \dot{X}$ exceeds the peak conservative torque (\ref{eq:Walker-peak-torque}) and the steady-state equilibrium is no longer possible. The motion of the domain wall becomes complex, combining non-steady precession of the azimuthal angle with oscillation and drift of position $X$. 

Fig.~\ref{fig:Walker-trajectory2} shows trajectories of a domain wall in configuration space $(X,\Phi)$ for several values of the field $h>h_c$. As in the steady-state regime, the trajectories start at $t=0$ at the minimum of the anisotropy energy $\Phi(0) = \pi/2$ and end at $t = 2T_c$, where $T_c = 2\pi/(\gamma h_c)$ is the Larmor period for the critical field. In contrast to the steady-state regime, increasing the field beyond $h_c$ leads to a reduction in the average velocity of the domain wall.   

\subsection{Oscillatory motion in a very strong field} 

In a very strong applied field, $h \gg h_c$, the domain wall moves in an oscillatory manner. To understand the nature of this motion, we may at first neglect Gilbert's damping and set $\alpha = 0$. The energy (\ref{eq:Walker-energy}) is then conserved,  
\begin{equation}
2 \sigma \mathcal M h X + U_0(\Phi) = \mathrm{const}.    
\end{equation}
This limits the range of motion for the domain wall's coordinate to
\begin{equation}
\Delta X = \frac{U_0(0) - U_0(\pi/2)}{2\mathcal M h}.    
\end{equation}
The azimuthal angle decreases steadily at the Larmor frequency (\ref{eq:Larmor-frequency}), 
\begin{equation}
\Phi(t) = \Phi(0) - \omega_L t. 
\end{equation}
The linear velocity can then be found from the balance of the gyroscopic and conservative torques, Eq.~(\ref{eq:Walker-balance-torques}), which yields 
\begin{equation}
\dot{X} = 
- \frac{\sigma \gamma \mathcal M \lambda_0}{8\pi} 
\sin{2\Phi(t)}.
\end{equation}
In the absence of damping, the velocity oscillates around zero average. 

Adding weak damping, $\alpha \ll 1$, breaks the conservation of energy. The fast oscillatory motion will lead to a gradual reduction of the energy with time. Averaged over an oscillation period, the domain-wall energy $U_0(\Phi)$ will remain constant over time, so the reduction must come from the Zeeman energy $2 \sigma \mathcal M h X$. That creates a slow drift on top of the oscillatory motion with an average velocity proportional to Gilbert's damping $\alpha$. In contrast, the steady-state velocity (\ref{eq:Walked-steady-state-velocity}) is proportional to $\alpha^{-1}$. The drift velocity above Walker's breakdown is thus much lower than the steady-state velocity below it. See Schryer and Walker \cite{Schryer1974} for further details and Beach \textit{et al.}~\cite{BeachNM2005} for experimental examples.

\subsection{D{\"o}ring's mass}

Let us return to the regime of a weak magnetic field. We will now investigate the transient dynamics of the wall before it reaches the steady state considered in Sec.~\ref{sec:Walker-steady-state}. 

The azimuthal angle of a weakly perturbed wall remains close to one of its equilibrium values, $\Phi = \pm \pi/2 + \delta \Phi$ with $\delta \Phi \ll 1$. The azimuthal angle is now a \emph{hard mode}. We will integrate it out to obtain the dynamics of the remaining \emph{soft mode} $X$. The elimination of the hard mode is also justified on practical grounds: it is easier for experimentalists to measure the position $X$ of a domain wall than to observe its azimuthal orientation $\Phi$. 

Near equilibrium, the conservative torque (\ref{eq:Walker-conservative-force-torque}) is approximately linear in the deviation $\delta \Phi$ from equilibrium, 
\begin{equation}
F_\Phi \sim - \kappa \, \delta \Phi,
\end{equation}
with the stiffness constant $\kappa = \mu^2 \epsilon_0$. We rewrite the equations of motion (\ref{eq:eom-Walker}) as follows: 
\begin{eqnarray}
X: && G_{X\Phi} \dot{\Phi} + F_X - D_{XX} \dot{X} = 0,
\nonumber\\
\Phi: && G_{\Phi X} \dot{X} - \kappa \, \delta \Phi = 0.
\end{eqnarray}
We have dropped the dissipative force $-D_{\Phi\Phi} \dot{\Phi}$ for the hard mode $\Phi$ because its motion is restricted. We will see later that this is a reasonable approximation. This simplification enables us to express the hard mode in terms of the soft mode, $\delta \Phi = G_{\Phi X} \dot{X}/\kappa$ and to eliminate it from the equation of motion for the soft mode, which acquires the familiar form of Newton's second law: 
\begin{equation}
- M \ddot{X} + F_X - D_{XX} \dot{X} = 0.
\label{eq:eom-Walker-X-only}
\end{equation}
Integrating out the hard mode $\Phi$ has endowed the soft mode $X$ with inertial mass 
\begin{equation}
M = G_{X\Phi}^2/\kappa
\label{eq:Doring-mass}
\end{equation}
and kinetic energy $M\dot{X}^2/2$. 

Equation (\ref{eq:eom-Walker-X-only}) has a characteristic relaxation time
\begin{equation}
\tau = \frac{M}{D_{XX}} 
= \frac{\tau_0}{\alpha \mu^2}
= \frac{4\pi}{\alpha \gamma \mathcal M},    
\end{equation}
during which the velocity of the domain wall reaches its terminal value $\dot{X} = F_X/D_{XX}$. We derived the effective inertial dynamics of the soft mode $X$ (\ref{eq:eom-Walker-X-only}) while neglecting the dissipative force $-D_{\Phi\Phi} \dot{\Phi}$ for the hard mode. To check the accuracy of that approximation, we extract the relaxation time from the full equations of motion (\ref{eq:eom-Walker}): 
\begin{equation}
\tau = \frac{G_{X\Phi}^2 + D_{XX} D_{\Phi\Phi}}{\kappa D_{XX}} 
= (1 + \alpha^2)\frac{M}{D_{XX}}.
\end{equation}
We can see now that dissipation of the hard mode gives a small correction $\mathcal O(\alpha^2)$ to the relaxation time, which justifies its neglect. 

As dissipative forces are not essential for the emergence of inertia, we may derive it in the Lagrangian framework. The low-energy Lagrangian of the position $X$ and azimuthal angle $\Phi = \pm \pi/2 + \delta \Phi$ can be written as
\begin{equation}
L(X,\delta \Phi) = - G_{X\Phi} \, \delta \Phi \, \dot{X} - \frac{\kappa \, \delta \Phi^2}{2} - U(X).
\label{eq:L-X-dPhi}
\end{equation}
The first term represents the geometric (Berry-phase) part of the Lagrangian (\ref{eq:L-soliton}) with the gauge choice $A_X = - G_{X\Phi} \, \delta \Phi$ and $A_\Phi = 0$. Minimization of the action $S = \int L(X,\delta \Phi)\, dt$ with respect to the hard variable $\delta \Phi$ yields the optimal $\delta \Phi = -G_{X\Phi} \dot{X}/\kappa$ for a given trajectory $X(t)$. Substituting this optimal $\delta \Phi(t)$ into the original Largangian (\ref{eq:L-X-dPhi}) yields a new Lagrangian for the soft mode $X$ alone,
\begin{equation}
L(X) = \frac{M \dot{X}^2}{2} - U(X),     
\end{equation}
with the D{\"o}ring mass (\ref{eq:Doring-mass}). It yields Eq.~(\ref{eq:eom-Walker-X-only}) without the dissipative force, which can be added via Rayleigh's dissipation function. 

The emergence of inertia in ferromagnetic solitons through the elimination of a hard mode was first noted by D{\"o}ring \cite{Doring:1948}. Inertia has been observed in the dynamics of domain walls \cite{Rado:1950, Saitoh2004} and skyrmions \cite{Makhfudz:2012, Buttner2015}. 

\section{Discussion}

The method of collective coordinates can be applied to more complex domain walls, e.g., composite \cite{Clarke2008} and extended \cite{Zhang:2018} ones, as well as to other types of solitons such as vortices \cite{Huber1982} and skyrmions \cite{Makhfudz:2012}. An extension to antiferromagnets \cite{Tveten:2013} shows that antiferromagnetic solitons generally possess kinetic energy and inertia. 

The method has been successful in describing the dynamics of magnetic solitons in classical spin systems. One potentially promising direction for further development is to generalize the collective-coordinate approach to quantum spin systems where quantum effects are prominent in the dynamics of solitons. 

The expert reader will note a glaring omission in our review: we chose not to cover the spin-transfer torque exerted by a spin-polarized electric current in a metallic ferromagnet \cite{Ralph:2008}. Spin-transfer torque has been treated in the framework of collective coordinates with the electrical current included as a background field \cite{Tserkovnyak:2008}. It has been recognized for some time that this approach is flawed and that a more sound treatment should include the elect
ric charge as one of the physical degrees of freedom, on the same footing as the collective coordinates of a magnetic soliton \cite{Dasgupta:2018}. A separate article in this issue by one of us addresses that problem \cite{Tchernyshyov:2022}.

\addcontentsline{toc}{section}{Acknowledgements}
\section*{Acknowledgements}
S.K.K. was supported by Brain Pool Plus Program through the National Research Foundation of Korea funded by the Ministry of Science and ICT (NRF-2020H1D3A2A03099291) and by the National Research Foundation of Korea funded by the Korea Government via the SRC Center for Quantum Coherence in Condensed Matter (NRF-2016R1A5A1008184). O.T. was supported by the U.S. Department of Energy, Office of Science, Basic Energy Sciences under Award No. DE-SC0019331.

\appendix

\section{}
\label{app:PT}

The Schr{\"o}dinger equation $\mathcal H_1 \psi = \epsilon \psi$ with a P{\"o}schl--Teller potential \cite{Poeschl1933}, 
\begin{equation}
\mathcal H_1 = - \frac{d^2}{dx^2} + 1 - 2 \mathrm{\,sech}^2{x},
\end{equation}
can be solved with the aid of supersymmetric quantum mechanics \cite{Cooper1995}. The Hamiltonian is factorized, $\mathcal H_1 = A^\dagger A$, in terms of non-Hermitian operators,
\begin{equation}
A = \frac{d}{dx} + \tanh{x},
\quad
A^\dagger = - \frac{d}{dx} + \tanh{x},
\end{equation}
with the commutator $[A,A^\dagger] = 2 \mathrm{\,sech}^2{x}$. 

The supersymmetric partner of $\mathcal H_1$ is 
\begin{equation}
\mathcal H_2 
	= A A^\dagger 
    = A^\dagger A + [A,A^\dagger]
    = \mathcal H_1 + 2 \mathrm{\,sech}^2{x}
    = - \frac{d^2}{dx^2} + 1,
\end{equation}
the Hamiltonian of a free non-relativistic particle. Its eigenstates are plane waves, $\psi_2(x) = e^{i k x}$, with energy eigenvalues $\epsilon_k = 1 + k^2$.

From each eigenstate of $\mathcal H_2$ we can construct an eigenstate of $\mathcal H_1$ with the same eigenvalue as follows. Suppose $A A^\dagger |\psi_2\rangle = \epsilon|\psi_2\rangle$. Make a new state $|\psi\rangle = A^\dagger |\psi_2\rangle$. It is an eigenstate of $\mathcal H_1$: 
\begin{eqnarray}
A^\dagger A |\psi\rangle 
	&=& A^\dagger A (A^\dagger |\psi_2\rangle)
    = A^\dagger (A A^\dagger |\psi_2\rangle)
\nonumber\\
    &=& A^\dagger (\epsilon |\psi_2\rangle)
    = \epsilon A^\dagger |\psi_2\rangle
    = \epsilon |\psi\rangle.
\end{eqnarray}
The norm of the new state is 
\begin{equation}
\langle \psi | \psi \rangle 
	= \langle \psi_2 | A A^\dagger | \psi_2 \rangle
    = \langle \psi_2 | \epsilon | \psi_2 \rangle
    = \epsilon \langle \psi_2 | \psi_2 \rangle.
\end{equation}
Therefore, a properly normalized eigenstate of $\mathcal H_1$ would be 
\begin{equation}
|\psi_1 \rangle 
	= \frac{A^\dagger}{\sqrt{\epsilon}} |\psi_2\rangle.
\end{equation}

We thus obtain eigenstates of $\mathcal H_1$ with eigenvalues $\epsilon_k = 1 + k^2$: 
\begin{equation}
\psi_k(x) 
	= \frac{1}{\sqrt{1+k^2}} 
		\left(
    		- \frac{d}{dx} + \tanh{x}
    	\right)
    	e^{ikx}
    = \frac{\tanh{x} - ik}{\sqrt{1+k^2}} e^{ikx}.
\label{eq:PT-free-states}
\end{equation}
Note that these states become plane waves far away from the potential well,  
\begin{equation}
\psi_k(x) 
	\sim \frac{\pm 1 - ik}{\sqrt{1+k^2}} e^{ikx}
    = e^{ikx + i \phi_\pm}
\mbox{ as }
x \to \pm \infty,
\end{equation}
where $\phi_{\pm} = -\pi/2 \pm \arctan{(1/k)}$. It is remarkable that the P{\"o}schl--Teller potential is reflectionless: a plane wave with wavenumber $k$ is fully transmitted and merely experiences a phase shift $\phi_+ - \phi_- = 2 \arctan{(1/k)}$.

Eigenstates of $\mathcal H_1$ have the same inner product as their supersymmetric partners, the plane waves:
\begin{equation}
\int dx \, \psi_{k'}^*(x) \psi_k(x) 
	= \int dx \, e^{-i k' x} e^{i k x}
	= 2\pi \delta(k - k').
\end{equation}

In addition to free states (\ref{eq:PT-free-states}), $\mathcal H_1$ has a bound state with exactly zero energy, $A^\dagger A \Psi_0 = 0$. This state is annihilated by the operator $A$, which yields a first-order differential equation 
\begin{equation}
\Psi_0'(x) + \tanh{x} \Psi_0(x) = 0.
\end{equation}
Its solution, properly normalized, is
\begin{equation}
\Psi_0(x) = 2^{-1/2} \mathrm{\,sech}{\,x}.
\end{equation}

\newpage 

\bibliography{solitons}

\providecommand{\newblock}{}
\begin{thebibliography}{10}
\expandafter\ifx\csname url\endcsname\relax
  \def\url#1{{\tt #1}}\fi
\expandafter\ifx\csname urlprefix\endcsname\relax\def\urlprefix{URL }\fi
\providecommand{\eprint}[2][]{\url{#2}}

\bibitem{Kosevich1990}
Kosevich A~M, Ivanov B~A and Kovalev A~S 1990 {\em Phys. Rep.\/} {\bf 194}
  117--238

\bibitem{bar2006dynamics}
Bar'yakhtar V~G, Chetkin M~V, Ivanov B~A and Gadetskii S~N 2006 {\em Dynamics
  of Topological Magnetic Solitons: Experiment and Theory\/} vol 129 (Springer)

\bibitem{Tretiakov2008}
Tretiakov O~A, Clarke D, Chern G~W, Bazaliy Y~B and Tchernyshyov O 2008 {\em
  Phys. Rev. Lett.\/} {\bf 100} 127204

\bibitem{ZangPRL2011}
Zang J, Mostovoy M, Han J~H and Nagaosa N 2011 {\em Phys. Rev. Lett.\/} {\bf
  107}(13) 136804

\bibitem{ThiavilleEPL2012}
Thiaville A, Rohart S, Ju{\'e} {\'E}, Cros V and Fert A 2012 {\em Euro. Phys.
  Lett.\/} {\bf 100} 57002

\bibitem{TataraPRB2020}
Tatara G and Otxoa~de Zuazola R~M 2020 {\em Phys. Rev. B\/} {\bf 101}(22)
  224425

\bibitem{FertNN2013}
Fert A, Cros V and Sampaio J 2013 {\em Nat. Nanotechnol.\/} {\bf 8} 152--156

\bibitem{ParkinNN2015}
Parkin S and Yang S~H 2015 {\em Nat. Nanotechnol.\/}  195--198

\bibitem{Altland-Simons}
Altland A and Simons B~D 2010 {\em Condensed Matter Field Theory\/} 2nd ed
  (Cambridge: Cambridge University Press)

\bibitem{Schryer1974}
Schryer N~L and Walker L~R 1974 {\em J. Appl. Phys.\/} {\bf 45} 5406--5421

\bibitem{Sakurai}
Sakurai J~J and Napolitano J~J 2020 {\em Modern Quantum Mechanics\/} 3rd ed
  (Pearson) ISBN 978-1108473224

\bibitem{Thomson:1879}
Thomson W and Tait P~G 1879 {\em Treatise on Natural Philosophy: Part 1\/}
  (Cambridge: Cambridge University Press) \S 345$^\mathrm{vi}$

\bibitem{Shapere:1989}
Shapere A and Wilczek F (eds) 1989 {\em Geometric Phases in Physics\/}
  (Singapore: World Scientific) ISBN 9971505991

\bibitem{Haldane1986}
Haldane F~D~M 1986 {\em Phys. Rev. Lett.\/} {\bf 57} 1488--1491

\bibitem{Gilbert2004}
Gilbert T~L 2004 {\em IEEE Trans. Mag.\/} {\bf 40} 3443--3449

\bibitem{Gonzalez2022}
Gonzalez-Meza R and Tchernyshyov O 2022 {\em J. Magn. Magn. Mater.\/} {\bf 562}
  169749

\bibitem{LL-I}
Landau L~D and Lifshitz E~M 1976 {\em Mechanics\/} 3rd ed ({\em Course of
  Theoretical Physics\/} vol~1) (New York: Butterworth-Heinemann) \S 25

\bibitem{Doring1948}
D\"{o}ring W 1948 {\em Z. Naturforsch.\/} {\bf 3A} 373--379

\bibitem{Z.Phys.61.206}
Bloch F 1930 {\em Zeitschrift f{\"u}r Physik\/} {\bf 61} 206

\bibitem{Yan:2011}
Yan P, Wang X~S and Wang X~R 2011 {\em Phys. Rev. Lett.\/} {\bf 107} 177207

\bibitem{Poeschl1933}
P{\"o}schl G and Teller E 1933 {\em Z. Phys.\/} {\bf 83} 143--151

\bibitem{Cooper1995}
Cooper F, Khare A and Sukhatme U 1995 {\em Phys. Rep.\/} {\bf 251} 267--385

\bibitem{Galkin2007}
Galkin A~Y and Ivanov B~A 2007 {\em J. Exp. Theor. Phys.\/} {\bf 104} 775--791
  ISSN 1063-7761

\bibitem{Takashima2016}
Takashima R, Ishizuka H and Balents L 2016 {\em Phys. Rev. B\/} {\bf 94} 134415

\bibitem{Goldstein2001}
Goldstein H, Poole Jr C~P and Safko J~L 2001 {\em Classical Mechanics\/} 3rd ed
  (New York: Pearson) {Chapter} 9.5

\bibitem{BeachNM2005}
Beach G~S~D, Nistor C, Knutson C, Tsoi M and Erskine J~L 2005 {\em Nat.
  Mater.\/} {\bf 4} 741 EP --

\bibitem{Doring:1948}
D{\"o}ring W 1948 {\em Z. Naturforsch.\/} {\bf 3a} 373--379

\bibitem{Rado:1950}
Rado G~T, Wright R~W and Emerson W~H 1950 {\em Phys. Rev.\/} {\bf 80} 273--280

\bibitem{Saitoh2004}
Saitoh E, Miyajima H, Yamaoka T and Tatara G 2004 {\em Nature\/} {\bf 432}
  203--206

\bibitem{Makhfudz:2012}
Makhfudz I, Kr\"uger B and Tchernyshyov O 2012 {\em Phys. Rev. Lett.\/} {\bf
  109} 217201

\bibitem{Buttner2015}
B{\"u}ttner F, Moutafis C, Schneider M, Kr{\"u}ger B, G{\"u}nther C~M, Geilhufe
  J, Schmising C~v~K, Mohanty J, Pfau B, Schaffert S, Bisig A, Foerster M,
  Schulz T, Vaz C~A~F, Franken J~H, Swagten H~J~M, Kl{\"a}ui M and Eisebitt S
  2015 {\em Nat. Phys.\/} {\bf 11} 225--228

\bibitem{Clarke2008}
Clarke D~J, Tretiakov O~A, Chern G~W, Bazaliy Y~B and Tchernyshyov O 2008 {\em
  Phys. Rev. B\/} {\bf 78} 134412

\bibitem{Zhang:2018}
Zhang S and Tchernyshyov O 2018 {\em Phys. Rev. B\/} {\bf 98} 104411

\bibitem{Huber1982}
Huber D~L 1982 {\em Phys. Rev. B\/} {\bf 26} 3758--3765

\bibitem{Tveten:2013}
Tveten E~G, Qaiumzadeh A, Tretiakov O~A and Brataas A 2013 {\em Phys. Rev.
  Lett.\/} {\bf 110} 127208

\bibitem{Ralph:2008}
Ralph D~C and Stiles M~D 2008 {\em J. Magn. Magn. Mater.\/} {\bf 320}
  1190--1216

\bibitem{Tserkovnyak:2008}
Tserkovnyak Y, Brataas A and Bauer G~E~W 2008 {\em J. Magn. Magn. Mater.\/}
  {\bf 320} 1282--1292

\bibitem{Dasgupta:2018}
Dasgupta S and Tchernyshyov O 2018 {\em Phys. Rev. B\/} {\bf 98} 224401

\bibitem{Tchernyshyov:2022}
Tchernyshyov O 2023 {\em J. Phys.: Condens. Matter\/} {\bf 51} 014001

\end{thebibliography}
\bibliographystyle{iopart-num}

\end{document}